\documentclass[preprint]{aastex}

\begin{document}

\title{High-Velocity Jets and Slowly Rotating Envelope in B335}

\author{Hsi-Wei Yen\altaffilmark{1,2}, Shigehisa Takakuwa\altaffilmark{2}, and Nagayoshi Ohashi\altaffilmark{2}}
\altaffiltext{1}{Institute of Astrophysics, National Taiwan University, Taipei 10617, Taiwan}
\altaffiltext{2}{Academia Sinica Institute of Astronomy and Astrophysics, P.O. Box 23-141, Taipei 10617, Taiwan} 

\begin{abstract}
We have performed detailed imaging and analyses 
of SMA observations in 230 GHz continuum, $^{12}$CO (2--1), $^{13}$CO (2--1),
and C$^{18}$O (2--1) emission toward B335,
an isolated and nearby ($\sim$ 150 pc) Bok globule with an embedded Class 0 source
($L_{\rm bol} \sim 1.5$ $L_{\sun}$). 
We report the first discover of  
high-velocity ($V_{\rm propagation} \sim$
160 km s$^{-1}$) $^{12}$CO (2--1) jets with a size of $\sim$ 900 AU $\times$ 1500 AU along the E-W direction in B335.
The estimated mass-loss rate ($\sim$ 2.3 $\times$ 10$^{-7}$ $M_{\sun}$ yr$^{-1}$)
and the momentum flux ($\sim$ 3.7 $\times$ 10$^{-5}$ $M_{\sun}$ yr$^{-1}$ km s$^{-1}$)
of the $^{12}$CO jets in B335 are one order of magnitude lower than those of
other $^{12}$CO jets in more luminous sources such as HH 211
($L_{\rm bol} \sim 3.6$ $L_{\sun}$) and HH 212 ($L_{\rm bol} \sim 14$ $L_{\sun}$).
The weaker jet activity
in B335 could be due to the lower active accretion onto the central protostar.
The C$^{18}$O emission shows a compact ($\sim$ 1500 AU) condensation associated
with the central protostar, and it likely traces the protostellar
envelope around B335, 
as in the case of the 230 GHz continuum emission.
The envelope exhibits a velocity gradient from the
east (blueshifted) to west (redshifted) that
can be interpreted as an infalling motion. The estimated central stellar mass,
the mass infalling rate, and the accretion luminosity are 0.04 $M_{\sun}$,
6.9 $\times$ 10$^{-6}$ $M_{\sun}$ yr$^{-1}$, and 2.1 $L_{\sun}$, respectively.
On the other hand, there is no clear velocity gradient perpendicular to the
outflow axis in the C$^{18}$O envelope, 
suggesting little envelope rotation on a hundred-AU scale.
The upper limits of the rotational velocity and specific angular momentum
were estimated to be 0.04 km s$^{-1}$ and 7.0 $\times$ 10$^{-5}$ km s$^{-1}$ pc
at a radius of 370 AU, respectively.
The specific angular
momentum and the inferred Keplerian radius ($\sim$ 6 AU)
in B335 are 1 - 2 orders of magnitude smaller than those in other more-evolved sources.
Possible scenarios to explain the lower specific angular momentum in B335 are discussed.
 
\end{abstract}

\keywords{circumstellar matter --- ISM: individual (B335) --- ISM : molecules --- stars : formation}

\section{Introduction}

Dense-gas condensations ($\geq$ 10$^{4-5}$ cm$^{-3}$) in dark molecular clouds are the sites
of low-mass star formation \citep{And00, Mye00}.
Previous millimeter interferometric observations have revealed rotating and infalling gas motions
in dense cores associated with known infrared sources, 
the so-called ``protostellar envelopes'' \citep{Oha96, Oha97a, Oha97b, Mom98}.
Bipolar molecular outflows associated with protostellar envelopes have also been observed  \citep{Bac99}. 
These molecular outflows are considered to
be ambient material entrained by the jet, 
ejected from the vicinity of the central protostar.
Eventually, the infall and the outflow
terminate, and a newly-born star surrounded by a circumstellar disk appears \citep{Shu87}.

Details of the physical processes in protostellar envelopes and outflows, however,
are still a matter of debate. 
For example, it is still unclear how the angular
momentum of the rotating motion in envelopes is transferred from large to small
radii (e.g., Goodman et al.; Ohashi et al. 1997b), 
and how
centrifugally-supported disks with radii of a few hundreds AU often observed around young stars 
(e.g., Guilloteau et al. 1999; Guilloteau \& Dutrey 1998; Qi et al. 2003) are formed in envelopes.
In order to address these questions, observations of the rotating motion in a
representative envelope, observed from large to small scales, and comparisons with other sources are required. 
On the other hand, the structure and kinematics of molecular outflows are different from source to source \citep{Lee00}. 
In several sources high-velocity ($>$ 100 km s$^{-1}$) collimated molecular jets
have been found, while in other sources only slow outflow shells with wide opening angles
are seen \citep{Bac99, Arc07}. The mechanisms which produce this variety of outflows,
and the relation of the mass ejection to the central mass accretion,
are still controversial.

B335 is an isolated Bok globule associated with an embedded far-infrared source (IRAS 19347+0727)
\citep{Kee80, Kee83}. The distance to B335 has been recently re-estimated to be $\sim$ 150 pc \citep{Stu08}, and in this paper, all the results referred from the literature
have been corrected using this new estimate. 
The central source
is a Class 0 source with a bolometric luminosity of 1.5 $L_{\sun}$ \citep{Stu08}
and a dust temperature of 31 K \citep{Cha93}.
$^{12}$CO (1-- 0) and $^{13}$CO (1-- 0) line observations of
B335 have unveiled the presence of a molecular outflow, 
both at
$\sim$ 0.2-pc \citep{Hir88,Cab88,Mor89} and 
$\sim$ 3000-AU scales
\citep{Hir92,Cha93}. The outflow extends along the E-W direction, 
and shows a conical shape with an opening angle of $\sim$ 45$\degr$
and an inclination angle of $\sim$ 10$\degr$ from the plane of the sky.
Along the outflow axis there are several HH objects (HH 119 A-F),
whose proper motions moving away from the central source
have been detected \citep{Rei92, Gal07}. 
The dynamic time scale of
the furthest HH object (HH 119 A) is $\sim$ 850 yr \citep{Rei92}, and the propagation
velocity reaches 140 -- 170 km s$^{-1}$ \citep{Gal07}.

Single-dish observations of B335 have found asymmetric profiles 
in optically-thick CS and H$_{2}$CO lines, implying the presence of infalling
motions in the envelope around B335 \citep{Zho93, Cho95}.
Direct interferometric imaging of the envelope around B335
in H$^{13}$CO$^{+}$ (1--0) and C$^{18}$O (1--0) at an
angular resolution of $\sim$ 6$\arcsec$
supports the presence of infalling motions at a 3000-AU
scales \citep{Sai99, Cha93},
although the interferometic imaging of
the CS (5--4) emission shows that the asymmetric line profiles are
influenced severely by contamination from the outflow \citep{Wil00}.
Recently, \cite{Cho07} has suggested that the asymmetric line profile
in the H$_{2}$CO line arises from both the outflow and the infalling
envelope. The envelope around B335 also exhibits a slow rotation at 
a radius of $\sim$ 20000 AU \citep{Fre87, Sai99} and $\sim$ 1000 AU
\citep{Sai99}. The density distribution in the envelope around B335
shows an $r^{-1.5}$ dependence between radii of 60 AU and 3900 AU,
while outside this region the radial dependence is $r^{-2}$
\citep{Har01, Har03a, Har03b}, which is consistent with the inside-out
collapse model \citep{Shu77}. B335 has also been observed in submillimeter
molecular lines \citep{Jor07, Tak07b}.

These results indicate that
B335 is a prototypical low-mass protostellar source suitable
for detailed studies. The observations at 1000-AU -- 40000-AU
scales described above have revealed the core, infalling and rotating
envelope, and the outflow. In the
present paper, 
we report detailed imaging and analyses of the Submillimeter Array (SMA)\footnotemark \
observations of the region within 1500 AU in B335
at an angular resolution of $\sim$ 4\arcsec, 
to study the inner part of the outflow and the envelope. 
Through the comparison with previous observations
of the envelope around B335 and those of other sources,
we will discuss how the angular momentum of the rotating motion in
envelopes is transfered. We will also compare the observed
properties of the outflow in B335 to those of other sources associated
with high-velocity molecular jets, and discuss the relation of the outflow
to the central mass-accretion processes.

\footnotetext{The Submillimeter Array (SMA) is a joint project
between the Smithsonian Astrophysical Observatory and the Academia
Sinica Institute of Astronomy and Astrophysics and is funded by
the Smithsonian Institute and the Academia Sinica.}

\section{Observation}
The present observations of B335 were made as a part of a large SMA project (PROSAC: J\o rgensen et al. 2007) on 2005 Jun 24 with the seven SMA antennas. Details of the SMA are described by \cite{Ho04}. The SMA is a double-sideband instrument with a 2 GHz bandwidth each. 
We observed 230 GHz continuum emission, $^{12}$CO (2--1; 230.5379700 GHz), $^{13}$CO (2--1; 220.3986765 GHz), and C$^{18}$O (2--1; 219.5603568 GHz) emission in B335 simultaneously. The pointing center was $\alpha$(J2000) = 19$^{h}$37$^{m}$00$\fs$89, $\delta$(J2000) = 7\arcdeg34\arcmin10$\farcs$0, and the Field of View was $\sim$ 55$\arcsec$ (8300 AU). 
The lengths of the projected baselines on the sky range from 5.5 $k\lambda$ to 53.5 $k\lambda$, and our observations were insensitive to structures more extended than $\sim$ 4500 AU at the 10\% level \citep{Wil94}.
The correlator configuration was set to assign 128 channels per one chunk with a 83.3 MHz bandwidth to the $^{12}$CO line, and 512 channels per chunk to the $^{13}$CO and C$^{18}$O lines, which results in a velocity resolution of 1.05, 0.28, and 0.28 km s$^{-1}$, respectively.

The passband calibrator was quasar 3C279, and the flux calibrator was Callisto. Quasar 1749+096 (1.9 Jy) and quasar 2145+067 (2.5 Jy) were observed as gain calibrators. The MIR software package was used to calibrate the data. The calibrated visibility data were Fourier-transformed and CLEANed with MIRIAD \citep{Sau95} to produce images. The observational parameters are summarized in Table 1. In order to improve the signal to noise ratio of the $^{12}$CO (2--1) data, we smoothed the data cube over 2 channels, and the noise level reduces to 100 mJy beam$^{-1}$.

\section{Results}

The images of B335 in the 1.3 mm continuum, $^{12}$CO (2--1), $^{13}$CO (2--1), and the C$^{18}$O (2--1) emission were first shown in the PROSAC paper \citep{Jor07}. In this paper, we present detailed results including velocity structures. 
Hereafter, the systemic velocity obtained from the single-dish results \citep{Hir91, Eva05}, 8.3 km s$^{-1}$, is adopted, and all the velocities are shown as the relative velocity ($\Delta$$V$) to this systemic velocity.

\subsection{1.3 mm Continuum Emission}

Figure 1 shows the 1.3 mm continuum image (contour) overlaid on the $^{12}$CO (2--1) moment 0 map (gray) in B335. The continuum emission shows an elongated structure along the N-S direction, plus two protrusions toward the N-W and S-W directions. The N-S elongation appears to be consistent with the previous single-dish result in the 1.3 mm continuum emission that shows an  
elongated feature with a size of $\sim$ 18000 AU $\times$ 13000 AU along the N-S direction \citep{Mot01}. 
As will be discussed in the next subsection, the $^{12}$CO emission most likely traces the molecular outflow along the E-W direction. Hence, the continuum emission is elongated almost perpendicularly to the E-W outflow, suggesting that the main component in the 1.3 mm continuum emission traces the circumstellar envelope around B335. On the other hand, the two protrusions abut upon the $^{12}$CO emission and delineate the rim of the outflow, which suggests that these components trace the wall of the cavity evacuated by the outflow.

By fitting a 2-dimensional Gaussian to the continuum image above the 6$\sigma$ level, which excludes the N-W and S-W protrusions, we obtained the peak position of $\alpha$(J2000) = 19$^{h}$37$^{m}$0$\fs$93, $\delta$(J2000) = 7\arcdeg34\arcmin09$\farcs$8, and hereafter we adopt this position as a position of the central protostellar source. The deconvolved size, position angle, and the total flux were estimated to be 4\farcs9 $\times$ 2\farcs3 (740 AU $\times$ 350 AU), 13\degr, and 0.18 Jy, respectively. The 1.3 mm continuum flux recovered with the SMA corresponds to $\sim$ 25 $\%$ of the integrated flux within a radius of 17\arcsec\ measured with the IRAM 30-m telescope \citep{Mot01}. We can estimate the mass of the main dusty component ($\equiv$ $M$) as 
\begin{equation}
M = \frac{F_{\nu}D^{2}} {\kappa_{\rm 230GHz} B(T_{\rm dust})},
\end{equation}
where $F_{\nu}$ is the total flux, $D$ is the distance to the source, $T_{\rm dust}$ is the dust temperature, and $B$ is the Planck function. On the assumption that the frequency ($\nu$) dependence of the dust mass opacity ($\tbond \kappa_{\nu}$) is $\kappa_{\nu} = 0.1 \times (\frac{\nu}{10^{12}})^{\beta}$ \citep{Bec90} with $\beta$ = 1.2 \citep{Cha93}, the mass opacity at 230 GHz ($\tbond \kappa_{\rm 230 GHz}$) was estimated to be  0.017 g cm$^{-2}$. 
The mass of the central compact component was estimated to be 0.027 $M_{\sun}$ at a dust temperature of 31 K \citep{Cha93}. 
Given the uncertainty of the missing flux, the estimated mass is consistent with the mass estimated from the 2.7 mm continuum emission by Owens Valley array observations (0.08 M$_{\sun}$; Chandler \& Sargent 1993).

\subsection{$^{12}$CO (2--1) Emission}

Figure 2 shows the distribution of the $^{12}$CO (2--1) emission in B335 integrated over the following three 
different velocity ranges; high velocity (HV; $\Delta$$V$ = -37.5 -- -18.5 \& 17.5 -- 36.5 km s$^{-1}$), middle velocity (MV; $\Delta$$V$ = -16.0 -- -8.7 \& 6.9 -- 15.3 km s$^{-1}$), and low velocity
(LV; $\Delta$$V$ = -5.8 -- -1.6 \& 0.6 -- 4.8 km s$^{-1}$). We detected the $^{12}$CO (2--1) emission at much wider velocity range ($\Delta$$V$ = -37.5 -- 36.5 km s$^{-1}$) than the single-dish result ($\Delta$$V$ = -5.3 -- 5.7 km s$^{-1}$) \citep{Hir91}. 
In the LV range, the blueshifted and redshifted emissions show a $V$-shaped geometry opening towards the east and west respectively, with its apex at the protostellar position. Similar but less significant blueshifted and redshifted $V$-shaped features are also seen on the other side, and hence the blueshifted and redshifted components are overlapped on each side. The maximum length, width, and opening angle of the $V$-shaped structures are $\sim$ 28$\arcsec$ (4200 AU), $\sim$ 24$\arcsec$ (3600 AU), and $\sim$ 60$\degr$, respectively. From low- to high-velocity, the morphology of the $^{12}$CO emission becomes more compact with less overlap between the blueshifted and redshifted components.
In the HV range, neither the blueshifted nor the redshifted emission shows the $V$-shaped morphology, but both show compact 
condensations with a size of $\sim$ 1500 AU $\times$ 900 AU 
located near the protostar.
The non-detection of the HV $^{12}$CO emission in the single-dish observation is probably due to the beam dilution effect since the HV $^{12}$CO emission is compact. 
The peak position of these HV components are located within $\sim$ 2$\arcsec$ ($\sim$ 300 AU) of the protostar. On the assumption of an outflow inclination angle of 10$\degr$\ from the plane of the sky \citep{Hir88}, 
the mean propagation velocity of the HV $^{12}$CO emission reaches $\sim$ 160 km s$^{-1}$, which is comparable to the velocity of the associated HH objects (140 -- 170 km s$^{-1}$) \citep{Gal07}, 
and the dynamic time is estimated to be $\sim$ 45 yr ($\tbond$ 1500 AU / 160 km s$^{-1}$).

Figure 3 presents Position-Velocity (P-V) diagrams of the $^{12}$CO emission along the E-W direction (P. A. = 90$\degr$) in B335. The P-V diagram at a higher velocity resolution ($2.1$ km s$^{-1}$; Figure 3 $left$) exhibits a spatially extended ($\sim$ 4200 AU $\times$ 3600 AU),
narrow-velocity ($\sim$ 10 km s$^{-1}$) feature 
of the LV component. The velocity structure of the LV component can be traced with a Hubble-like velocity law ($v = C \times r$) in a geometrically-thin conical outflow shell with an opening angle of 60\degr\ and an inclination angle of 10$\degr$, where the coefficient $C$ is found to be 1.2 $\times$ 10$^{-3}$ km s$^{-1}$  AU$^{-1}$ (orange dotted cross in Figure 3). These results suggest that the LV $^{12}$CO emission delineates the rim of the E-W outflow with a conical shape, and that the outflow axis is slightly tilted with its eastern part on the near-side to us. 
Previous single-dish observations of B335 in $^{12}$CO (2--1 and 1--0) have found that there is an 
$\sim$ 0.2-pc
conical-shaped outflow along the E-W direction with an opening angle of $\sim$ 45$\degr$, 
and that the the eastern outflow axis is tilted toward us by $\sim$ 10$\degr$ from the plane of the sky (e.g., Hirano et al. 1988 \& 1991). The Spitzer IRAC image in B335 shows a reflection nebula with its eastern lobe on the near side \citep{Stu08}. The single-dish and Spitzer results are consistent with our results, and the LV $^{12}$CO (2--1) emission observed with the SMA most likely traces the basement of the extended outflow in B335.

On the other hand, the P-V diagram at a smoothed velocity resolution ($4.2$ km s$^{-1}$; Figure 3 $right$) with a better sensitivity shows 
that the kinematics of the HV $^{12}$CO is distinct from that of the LV component; 
the HV component exhibits a much wider velocity width ($\sim$ 20 km s$^{-1}$) than the LV component though the spatial extent of the HV component ($\sim$ 1500 AU $\times$ 900 AU) is much smaller than that of the LV component. Hence, the HV $^{12}$CO emission likely traces distinct outflow components from the LV outflow shell.
On the assumption of LTE conditions and optically-thin $^{12}$CO emission with an abundance of 10$^{-4}$ \citep{Fre87, Luc98}, and an excitation temperature of 50 K \citep{Lee07b}, 
the column density at the peaks was estimated to be 6.7 $\times$ 10$^{19}$ cm$^{-2}$, and 
the gas masses of the redshifted and blueshifted HV components were estimated to be 2.3 $\times$ 10$^{-5}$ and 2.0 $\times$ 10$^{-5}$ $M_{\sun}$, respectively. 
While the SMA $^{12}$CO observations miss 90$\%$ of the total flux observed with JCMT \citep{Hir91} around the systemic velocity, the missing flux decreases to $\leq$ 30\% at a relative velocity of more than 3 km s$^{-1}$. 
The flux from the HV component ($\Delta$$V$ $\ga$ 18 km s$^{-1}$), which is compact and not detected by JCMT, is mostly recovered with the present SMA observation. 
We will discuss the origin of the HV components in $\S$4.1 in more detail.

\subsection{$^{13}$CO (2--1) Emission}

Figure 4 presents moment 0 maps of the blueshifted ($\Delta$$V$ = -2.1 -- 0.0 km s$^{-1}$) and redshifted ($\Delta$$V$ = 0.2 -- 1.9 km s$^{-1}$) $^{13}$CO (2 -- 1) emission in B335. The total velocity range of the $^{13}$CO emission ($\Delta$$V$ = -2.1 -- 1.9 km s$^{-1}$) is within the velocity range of the $^{12}$CO LV component. 
We note that 
the absence of the $^{13}$CO counterpart to the $^{12}$CO HV component is likely due to the insufficient sensitivity of the observations.
Both the blue- and redshifted $^{13}$CO emission consists of central compact 
components with a size of $\sim$ 1000 AU,
and outer extensions, while the blueshifted emission is more intense than the redshifted emission. 
There are two redshifted ``arms'' extending towards the S-E and S-W directions and a blueshifted arm towards the S-E direction. 
These outer extensions, as well as the slightly extensions to the N-E and N-W of the protostar at both blueshifted and redshifted velocities, form structures similar to the $V$-shaped geometry seen in the LV $^{12}$CO emission. 
The interferometric image of B335 in $^{13}$CO (1--0) \citep{Cha93} emission shows a similar morphology to that in the $^{13}$CO (2--1) emission, but the $V$-shaped geometry is less clear in the lower excitation line. 
The $^{13}$CO (2--1) P-V diagram along the outflow axis shows a velocity structure that can be explained as a Hubble-like velocity law in a geometrically-thin conical outflow shell on a size scale of $\sim$ 1500 AU, as shown as the solid lines in Figure 5. A similar velocity structure was also seen in the $^{12}$CO P-V diagram along the outflow axis, but on a size scale of $\sim$ 2500 AU.
The more intense blueshifted emission compared to the redshifted emission could be due to less absorption at the front side of the outflow shell than at the back. 
In the outflow-envelope configuration, the redshifted outflow lobe is located behind 
with respect to the blueshifted lobe, and hence will suffer more
from absorption by foreground material.

Assuming a $^{13}$CO abundance of 1.7 $\times$ 10$^{-6}$ \citep{Fre87}, an excitation temperature of 20 K, and LTE and optically-thin conditions, the gas masses traced by the blueshifted and the redshifted $^{13}$CO emission were estimated to be 1.9 $\times$ 10$^{-3}$ $M_{\sun}$ and 3.2 $\times$ 10$^{-3}$ $M_{\sun}$, respectively. 
The momenta of the blueshifted and redshifted $^{13}$CO components were estimated to be 2.3 $\times$ 10$^{-3}$ and 2.1 $\times$ 10$^{-3}$ $M_{\sun}$ km s$^{-1}$, respectively, using the velocity channel maps ($P_{^{13}\rm CO}$ = $\sum$ $M_{\rm channel}$ $\times$ $V_{\rm channel}$). 
Here, we assumed that the $^{13}$CO emission traces the same geometrically-thin conical shell as the LV $^{12}$CO component, and we corrected for the inclination to estimate the propagation velocity. 
The estimated masses and momenta should be considered as lower limits, since only $\sim$ 40$\%$ of the total $^{13}$CO (2--1) flux obtained with JCMT \citep{Hir91} is recovered with the present SMA observation.

\subsection{C$^{18}$O (2--1) Emission}

Figure 6 presents the moment 0 map of the C$^{18}$O (2--1) emission integrated from $\Delta$$V$ = -0.9 km s$^{-1}$ to 0.8 km s$^{-1}$ in B335. In contrast to the $^{12}$CO and $^{13}$CO emission, 
the C$^{18}$O emission shows a 
$\sim$ 1500-AU
condensation 
with a single peak slightly offset 
from the protostellar position by $\sim$ 1\arcsec. 
The condensation shows an almost spherical structure with N-E, S-E, and weak S-W extensions, which is consistent with the C$^{18}$O (1--0) map obtained using the Owens Valley millimeter array \citep{Cha93}. 
Single-dish observations of B335 in the C$^{18}$O (1--0) line show a 
$\sim$ 36000 AU $\times$ 32000 AU
 envelope with a mass of 2.4 $M_{\sun}$ around the protostar with an also almost spherical structure \citep{Sai99}. 

Figure 7 shows the velocity channel maps of the C$^{18}$O (2--1) emission overlaid with the 1.3 mm continuum emission (see Fig. 1). At $\Delta$$V$ = -0.9 km s$^{-1}$, weak C$^{18}$O emission elongated along the N-S direction was detected to the east of the protostellar position. At $\Delta$$V$ = -0.6 km s$^{-1}$, C$^{18}$O emission is elongated along the outflow axis (P. A. = 90\degr), and at $\Delta$$V$ = -0.3 km s$^{-1}$ C$^{18}$O emission is elongated toward the SE. 
Around the systemic velocity ($\Delta$$V$ = 0.0 -- 0.2 km s$^{-1}$), C$^{18}$O emission is elongated perpendicularly to the outflow axis, and its distribution is similar to that of the 1.3 mm dust emission. At a redshifted velocity of $\Delta$$V$ = 0.5 km s$^{-1}$, C$^{18}$O emission shows a compact blob as well as 
a weak SW extension with its peak position slightly shifted to the west of the protostar. In the channel maps, the peaks of the C$^{18}$O emission shift from east to west of the protostar as the velocity changes from blueshifted to redshifted.

Figure 8 shows P-V diagrams of the C$^{18}$O emission along ($left$ $panel$) and across ($right$ $panel$) the outflow axis passing through the central protostar. Along the outflow axis there appears a velocity gradient: there are blueshifted and redshifted C$^{18}$O emission peaks at the east and west of the protostar, respectively. This velocity gradient could be due to the outflow. 
The P-V diagram, however, does not show the $X$-shaped velocity structure seen in the $^{12}$CO and $^{13}$CO emission. 
In fact, the model explaining the $^{12}$CO emission cannot be applied to the C$^{18}$O emission 
(bold solid lines in Figure 8 $left$). 
On the other hand, there is no clear velocity gradient seen in the C$^{18}$O P-V diagram perpendicular to the outflow axis. We will discuss the origin  and kinematics of the C$^{18}$O emission in $\S$4.2.

The brightness ratio between the C$^{18}$O (2--1) and C$^{17}$O (2--1) emission observed with the CSO indicates that the C$^{18}$O (2--1) emission is optically thin ($\tau_{\rm C^{18}O\ (2-1)} \sim 0.8$) \citep{Eva05}. 
Assuming LTE conditions, a C$^{18}$O abundance of 3 $\times$ 10$^{-7}$ \citep{Fre87}, and an excitation temperature of 30 K \citep{Cha93}, 
we estimated the total gas mass of the C$^{18}$O condensation to be 5.2 $\times$ 10$^{-3}$ $M_{\sun}$ from the total C$^{18}$O integrated intensity 
($\sim$ 9400 K km s$^{-1}$ over $\sim$ 140 arcsec$^{2}$). 
The derived gas mass is six times smaller than the mass estimated from the 1.3 mm dust continuum emission ($\sim$ 0.03 $M_{\sun}$), 
although the extent of the C$^{18}$O emission is approximately two times larger than that of the 1.3 mm dust continuum emission. 
This suggests that the C$^{18}$O abundance may be approximately one order of magnitude smaller that the value we assumed above, 
if the molecular gas and dust are well mixed and both the C$^{18}$O and 1.3 mm emission trace the same structure.
The lower C$^{18}$O abundance in the central region of the envelope could be due to the depletion of CO molecules onto dust grains. 
In fact, a similar degree of CO depletion has been suggested by modeling the single-dish line profiles of B335 \citep{Eva05}. Although the missing flux of the SMA C$^{18}$O data is estimated to be 80\% comparing to the single-dish flux of the C$^{18}$O (2--1) emission \citep{Eva05},
the above discussion on the lower C$^{18}$O abundance is still valid because the C$^{18}$O and 1.3 mm dust continuum emissions miss a similar amount of flux. 

\section{Discussion}
\subsection{High-velocity $^{12}$CO Component}

Along with the $V$-shaped conical outflow shell, we have discovered compact ($\sim$ 1500 AU $\times$ 900 AU) high-velocity ($V_{\rm propagation}$ $\sim$ 160 km s$^{-1}$) $^{12}$CO (2--1) components in B335 with our SMA observations. Such an outflow configuration with collimated high-velocity $^{12}$CO components plus low-velocity $^{12}$CO outflow shells is also seen in other low-mass protostellar sources associated with molecular jets such as HH 211 \citep{Lee06}, HH 212 \citep{Gue99}, and L1448-mm \citep{Bac95}, and has been explained by the jet-driven bow-shock model \citep{Bac95, Lee06} and the wind-driven model \citep{Sha07}. In Table 2 we compare the kinematical properties of the compact high-velocity $^{12}$CO (2--1) components in B335 
to those of the three $^{12}$CO jets mentioned above. The line width and the propagation velocity of the compact high-velocity $^{12}$CO  components in B335 are comparable to those in the other sources. 
In addition, in B335 there are several HH objects (HH 119 A-F) aligned along the outflow axis \citep{Gal07}, which should trace bow shocks at the leading heads of the episodic mass ejection, and the propagation velocity of the HV $^{12}$CO components is consistent with that of these HH objects ($\sim$ 160 km s$^{-1}$). These facts suggest that the high-velocity $^{12}$CO components found in B335 are most likely molecular jets and counterparts of the high-velocity $^{12}$CO jets seen in HH 211, HH 212, and L1448-mm.

We estimated the activity of the high-velocity $^{12}$CO jets 
in B335 with the method adopted by \cite{Lee07a, Lee07b} for the HH 211 and HH 212 jets. On the assumption that the transverse width of the $^{12}$CO jets in B335 is 300 AU as in the case of HH 211 \citep{Lee07b}, the volume gas density in the $^{12}$CO jets ($\tbond n_{\rm jet}$) was estimated to be $\sim$ 1.5 $\times$ 10$^{4}$ cm$^{-3}$ (from the column density of $\sim$ 6.7 $\times$ 10$^{19}$ cm$^{-2}$; see $\S$3.2). By assuming that the jet morphology is cylindrical, the mass-loss rate ($\dot{M}_{\rm loss}$) can be derived as  
\begin{equation}
\dot{M}_{\rm loss} = r^{2}\pi V_{\rm jet}n_{\rm jet}\mu,
\end{equation}
where $r$, $V_{\rm jet}$, and $n_{\rm jet}$ represent radius, propagation velocity, and the volume gas density in jets, respectively, and $\mu$ is the mean molecular weight. Then the mass-loss rate was estimated\footnotemark\ to be 2.3 $\times$ 10$^{-7}$ $M_{\sun}$ yr$^{-1}$, and the momentum flux ($ \tbond F = \dot{M}_{\rm loss} \times v_{\rm jet}$) was estimated to be 3.6 $\times$ 10$^{-5}$ $M_{\sun}$ yr$^{-1}$ km s$^{-1}$. 
In Table 3, we compare the estimated jet activity in B335 to those in HH 211 and HH 212. 
The density, the mass-loss rate, and the momentum flux in B335 are one order of magnitude lower than those in HH 212 and HH 211. 
In B335, thermal SiO emission, which is considered to be an excellent tracer of bow shocks in protostellar jets \citep{Bac96}, was not detected \citep{Jor07}, while intense SiO emission was found in HH 211 (e.g., Lee et al. 2007b), HH 212 (e.g., Lee et al. 2007a), and L1448-mm (e.g., Girart et al. 2001). 
Moreover, through high--$J$ $^{12}$CO line observations in far-infrared, the temperature of the outflow in B335 
was estimated to be $\sim$ 350 K with a LVG model, 
and it
is lower than that in the other protostellar jets, such as L1448-mm ($\sim$ 1200 K) and HH 211 (350--950 K) \citep{Gia01}. 
In the case of HH 211, HH 212 and L1448-mm, the $^{12}$CO jets conform to a chain of knots between the two successive extended bow shocks, while in B335 only one high-velocity $^{12}$CO knot located close to the central protostar ($\sim$ 1500 AU) is found. These results suggest that the jet phonemena in B335 are less active than those in the other sources.

\footnotetext{If we estimate $\dot{M}_{\rm loss}$ by $M_{\rm jet} / T_{\rm dynamic}$, the value becomes three times larger.}

In Table 4, we compare the protostellar properties of B335 to those of the other driving sources of the jets. It is clear that the bolometric luminosity of B335 is lower than that of HH 212, 211, and L1448-mm, although their bolometric temperatures are similar. The ratio between the bolometric luminosity and the central stellar mass, which should be proportional to the mass accretion rate in the central accretion disk, is more than 2-3 times lower in B335 than in the other sources. Hence, the lower mass-loss rate and the weaker jet activity in B335 compared to the other sources are likely to be linked with the lower mass accretion rate in B335. We suggest that the jet activity is closely related to the properties of the central accretion process.

\subsection{The Origin and Kinematics of the C$^{18}$O Emission}
As shown in Figure 7, the C$^{18}$O emission at $\Delta$$V$ = -0.6 and -0.3 km s$^{-1}$ shows elongation toward the east and S-E directions, respectively, which is similar to that of the outflow observed in $^{12}$CO (2--1) and $^{13}$CO (2--1) emissions. On the other hand, at around the systemic velocity ($\Delta$$V$ = 0.0 and 0.2 km s$^{-1}$), the C$^{18}$O emission is elongated perpendicularly to the outflow axis. In addition, there is a velocity gradient in the C$^{18}$O emission along the outflow axis at around the systemic velocity, 
while no clear velocity gradient is seen across the outflow axis.
In order to study the origin and kinematics of the C$^{18}$O emission, 
in Figure 9, we compare the C$^{18}$O emission, integrated in four different velocity ranges, with the $^{13}$CO emission integrated in the same velocity ranges and with the 1.3 mm dust continuum emission. 
At around the systemic velocity ($\Delta$$V$ = 0.0 and 0.2 km s$^{-1}$), both the blueshifted and redshifted C$^{18}$O emission shows clear elongation perpendicular to the outflow axis and 
resembles the dust continuum emission in morphology. 
In contrast, the $^{13}$CO emission at the same velocity is elongated along the outflow axis, and is likely to trace the outflow.  The highly blueshifted C$^{18}$O emission ($\Delta$$V$ = -0.9 -- -0.3 km s$^{-1}$) shows elongation toward the SE with a slight extension toward the NE, and this morphology is similar to that of the $^{13}$CO emission at the same velocity. The highly redshifted C$^{18}$O emission ($\Delta$$V$ = 0.5 -- 0.8 km s$^{-1}$) shows a central condensation elongated perpendicularly to the outflow axis with a weak extension toward the SW. 
Although the C$^{18}$O emission shows an extension similar to the $^{13}$CO outflow shells,
the overall structure of the C$^{18}$O central condensation is not similar to that of the $^{13}$CO outflow at the same velocity range. 
Therefore, the C$^{18}$O emission (at least at around the systemic velocity and probably at the highly redshifted velocity) traces structures different from the outflow traced by the $^{13}$CO emission, and  most likely traces the flattened molecular envelope 
perpendicular to the outflow axis. 
Such a flattened molecular envelope has often been observed around low-mass protostars (e.g., Ohashi et al. 1997a). 
Because the molecular outflow associated with B335 is aligned the plane of sky closely, it is naturally expected 
for the flattened envelope to have an almost edge-on configuration
and to show an elongated structure seen in channel maps.

Since the blueshifted and redshifted outflow emissions arise mostly from the east and west sides of the protostar, respectively, 
the eastern lobe of the outflow is tilted toward us from the plane of the sky. Hence, the eastern part of the envelope is tilted away from us, while the western part of the envelope is tilted toward us from the plane of the sky.
In this configuration, the eastern part of the envelope is the far side, while the western side is the near side. 
The flattened envelope shows a velocity gradient along its minor axis with the blueshifted emission at the far side and the redshifted emission at the near side. This suggests that the flattened envelope has infalling motions toward the central protostar. 
In Figure 10, we show a schematic picture of the configuration of the outflow and the infalling flattened envelope described above. 
From the peak offset between the channels at $\Delta$$V$ = 0.0 and 0.2 km s$^{-1}$, where the C$^{18}$O emission most likely traces the flattened envelope, the velocity gradient along the minor axis was measured to be 3.7 $\times$ 10$^{-3}$ km s$^{-1}$ AU$^{-1}$. From this velocity gradient, the infall velocity was estimated to be 0.28 km s$^{-1}$ at a radius of 440 AU on the assumption of an inclination angle of 10\degr. 
This infall velocity yields a central stellar mass of 0.02 $M_{\sun}$ in the case of the free-fall motion. 

On the other hand, there is no detectable velocity gradient between these two channels ($\Delta$$V$ = 0.0 and 0.2 km s$^{-1}$) across the outflow axis, suggesting no detectable rotation in the flattened envelope on a hundred-AU scale. 
Although there is no detectable rotation in the envelope on a hundred-AU scale, the flattened envelope could be produced by magnetic field (e.g., Galli \& Shu 1993). In addition, an outflow can sweep away a part of the material in the envelope along the outflow axis, making the shape of the envelope flattened. Two Class 0 sources, NGC 1333 IRAS 2A \citep{Bri09} and IRAS 16293-2422 \cite{Tak07a}, show similar cases; their innermost envelopes showing disk-like structures also have no detectable rotation on a hundred-AU scale.

In order to 
study the kinematics of the envelope in more detail, 
we constructed a simple model of a geometrically-thin infalling and rotating envelope with 
a Gaussian intensity distribution and compared it with the observations.
Note that even though the actual envelope has a thickness, we use a model without thickness to make the model simpler.
Although the Gaussian intensity distribution is an arbitrary choice, the choice of the intensity distribution does not affect main velocity structures. 
The radius of the model envelope was set to be 370 AU based on the semi-major axis of the 1.3 mm dust continuum emission (see $\S$3.1).
The inclination angle of the outflow \citep{Hir88} was adopted as that of the envelope. 
The radial motion of the model envelope due to the dynamical infall is described as $v_{r}(r) = \sqrt{\frac{2GM_{*}}{r}}$, where $M_{*}$ is the mass of the central star,
while its angular motion due to rotation is described as $v_{\phi}(r) \propto r^{-1}$ because of the angular momentum conservation. 
Based on the measurement of the infall velocity using the channel maps described above, 
the stellar mass was initially set to be 0.02 $M_{\sun}$. The rotational velocity was set to be 0.04 km s$^{-1}$ at $r$ = 370 AU, which corresponds to the detection limit, because there is no detectable rotation.
Then we generated synthesized images of the model envelope with the same uv-sampling as our SMA observations.

For the comparison between the model and the observations, we produced P-V diagrams from both the model and observations as shown in Figure 11.
Green contours in Figure 11 $(a)$ and $(g)$ show the P-V diagrams derived from the model. 
This simple model can reproduce the main feature of the C$^{18}$O P-V diagram both along the major and minor axis. 
We note that there is a velocity difference in the redshifted peak between the model and the observations. 
In order to match the model redshifted peak with the observed peak, we need to adopt a higher stellar mass in the model. In the case of a model with a stellar mass of 0.04 $M_{\sun}$ as shown in Fig. 11 $(b)$ and $(h)$, 
the model redshifted peak better matches the observed peak. 
If we adopt an even higher stellar mass (0.08 $M_{\sun}$; Fig. 11 $(c)$ and $(i)$), the model blueshifted peak would be offset from the observed peak due to the larger line width. 
On the other hand, if we set a higher rotational velocity such as $v_{\phi}$ = 0.16 km s$^{-1}$ at $r$ = 370 AU in the models (Fig. 11 $(j)$--$(l)$), there is a clear velocity gradient in the model P-V diagrams across the outflow axis, different from the observations which show no detectable velocity gradient across the outflow axis. Therefore, the model with a stellar mass of 0.04 $M_{\sun}$ and a rotational velocity of 0.04 km s$^{-1}$ at a radius of 370 AU provides a P-V diagram that matches the observations best.
We note that there is a slight positional shift in the blueshifted peak between the model and observations in the P-V diagram across the outflow axis.
Although a higher inclination angle could provide a better match of the model blueshifted peak with the observed peak, the difference may be due to contamination of outflowing motions in the observations since part of the blueshifted emission may arise from the outflow.

\subsection{Infalling Motion in the Envelope}
From our simple model of the infalling envelope (see $\S$4.2), the infalling velocity in the envelope around B335 was estimated to be 0.31--0.44 km s$^{-1}$ at a radius of 370 AU, which corresponds to a central stellar mass ($\equiv$ $M_{*}$) of 0.02--0.04 $M_{\sun}$. With an envelope mass ($\equiv$ $M_{\rm env}$) of 0.027 $M_{\sun}$ derived from the 1.3 mm continuum emission, the mass infalling rate
($\dot{M}_{\rm inf} = M_{\rm env}V_{\rm inf} / R_{\rm inf}$) and the accretion luminosity ($L_{\rm acc}$ = $GM_{*}\dot{M}_{\rm inf} / 4R_{\sun}$) were estimated to be 4.8--6.9 $\times$ 10$^{-6}$ $M_{\sun}$ yr$^{-1}$ and 0.7--2.1 $L_{\sun}$, respectively, where 4 R$_{\sun}$ is the radius of the protostar \citep{Sta80}. Hence, in B335, the mass outflow rate of the $^{12}$CO jets ($\dot{M}_{\rm out}$ = 3.1 $\times$ 10$^{-7}$ $M_{\sun}$ yr$^{-1}$) is 4--7 $\%$ of the mass infalling rate, 
which is slightly smaller than the results in HH 212 \citep{Lee07a} and HH 211 \citep{Lee07b}, and the accretion luminosity is comparable to or smaller than the bolometric luminosity (1.5 $L_{\sun}$).

In B335, the knotty distribution of the associated HH objects along the outflow axis likely suggests episodic jet ejection, and the high-velocity $^{12}$CO jet components found in our SMA observations are likely to represent the latest ejection event. Episodic jet ejections have been found in a number of protostellar sources, and are often linked to a sudden increase in the mass accretion \citep{Arc07}. In fact, \cite{Dun08} and \cite{Eno09} have revealed that many protostellar sources show one order of magnitude lower bolometric luminosity than the accretion luminosity predicted from the steady mass accretion model, and have proposed that the accretion is episodic and sources with lower bolometric luminosity than  model predictions are probably in the quiescent stage. Direct observational comparisons between the accretion luminosity derived from the observed infalling motion in the envelopes and the bolometric luminosity also suggest non-steady mass accretion. In HL Tau \citep{Lin94, Hay93}, L1551 IRS 5 \citep{Oha96, Sai96}, and in IRAS 16293-2422 \citep{Tak07a}, the estimated accretion luminosities are an order of magnitude higher than the bolometric luminosity, and in HH 212 \citep{Zin98, Lee06} the accretion luminosity  ($\sim$ 7 $L_{\sun}$) is lower than the bolometric luminosity (14 $L_{\sun}$). This mismatch could be reconciled by a picture similar to the FU-Ori phenomenon \citep{Har96} and episodic mass accretion. The protostar is surrounded by a disk, and outside of the disk there exists an infalling envelope. The material in the envelope keeps infalling onto the disk but not directly onto the surface of the protostar. In the "non-active" phase, most material in the disk does not accrete onto the surface of the protostar, and the jet ejection is also quiescent. Hence, the accretion luminosity derived from the mass infalling rate in the infalling envelope could be higher than the bolometric luminosity, as in the case of HL Tau, L1551 IRS5, and IRAS 16293-2422. When the disk becomes massive and unstable, the material accumulated in the disk starts falling onto the surface of the protostar, and then the powerful mass ejection also occurs \citep{Har96}. This is the ``active'' phase, when the accretion luminosity estimated from the outer infalling envelope could be lower than the ``real'' accretion luminosity, as in the case of HH 212 associated with the clear high-velocity jets.
Our detection of high-velocity molecular jets with a short dynamic time ($\sim$ 45 yr), and a possibly lower accretion luminosity than the bolometric luminosity in B335, imply a recent burst of mass ejection and accretion. 
In addition, the [OI] line emission has been detected around the protostar in B335, which could suggest the presence of shocks due to a recent ejection of the jets \citep{Nis99}. These results are consistent with the idea that B335 is in an "active" accretion phase. 

\subsection{Non-Conserved Angular Momentum in B335}

The infalling envelope traced by C$^{18}$O (2--1) emission in B335 does not show any clear velocity gradient perpendicular to the outflow axis, suggesting an absence of rotational motion in the envelope at the 300-AU scale. 
The upper limit of the specific angular momentum was estimated to be 7 $\times$ 10$^{-5}$ km s$^{-1}$ pc (see $\S$4.2), 
which corresponds to a rotational velocity of 0.04 km s$^{-1}$ at a radius of 370 AU. 
Since the material within this radius is considered to be dynamically infalling (see $\S$4.3), 
the specific angular momentum of the material within this radius is supposed to be conserved. 
If this is the case, 
the centrifugal force of the rotation becomes balanced with the gravitational force due to the central protostar with a mass of 0.04 M$_{\odot}$ at a radius of $\sim$ 6 AU, which can be considered an upper limit for the radius of the Keplerian rotating disk. The upper limit of the specific angular momentum at the small scale ($\sim$ 7 $\times$ 10$^{-5}$ km s$^{-1}$ pc) is, however, much lower 
than the measured specific angular momenta at radii of 1000 AU ($\sim$ 5.4 $\times$ 10$^{-4}$ km s$^{-1}$ pc) and 20000 AU ($\sim$ 4.6 $\times$ 10$^{-3}$ km s$^{-1}$ pc) \citep{Sai99}. If the angular momentum is conserved
from large to small scales, the material falling from a radius of 20000 AU should rotate at a velocity of 2.8 km s$^{-1}$ at a radius of 370 AU, which is 70 times larger than the upper limit of the rotational velocity estimated using the present observations. These results show that the rotational motion in the envelope around B335 is spinning-down toward the inner radii.

This decrease in the specific angular momentum from large to small scales has also been found in other protostellar sources and NH$_{3}$ cores by \cite{Oha97b} and \cite{Goo93}.
The specific angular momentum in B335 at a 
radius $>$ 20000 AU is similar to that of the NH$_{3}$ cores,
however, the specific angular momentum on the hundred-AU scale in B335 is one order of magnitude smaller than that in other protostellar sources (see Table 5). This lower specific angular momentum on the small scale could be explained by evolutional effects as follows.
A study of velocity gradients in NH$_{3}$ cores by \cite{Goo93} has shown that the specific angular momentum is larger at a larger radius (i.e., $j \propto r^{1.6}$). If B335 is in an early phase of the inside-out collapse of such a dense core, the material in an outer region with a larger angular momentum has not yet fallen into the central region, 
and only the material at an inner radius with a smaller angular momentum has fallen in dynamically. 
Hence, the specific angular momentum on the hundred-AU scale in B335 could be still small.

To test this scenario, we estimate the size of the dynamical-infalling region 
and the maximum amount of the specific angular momentum carried in by the dynamical infall. 
Based on single-dish observations of the envelope around B335 in C$^{18}$O (1--0) emission by \cite{Sai99} and the core rotation profile found by \cite{Goo93}, we assume that the initial condition of the core in B335 is a 
sphere with a rotation profile $\propto$ $r^{0.6}$, a density profile $\propto$ $r^{-2}$, a total mass of 2.4 $M_{\sun}$, and a core radius of 20000 AU. 
Our SMA observation shows that the total mass of the 
region within a radius of 370 AU is $\sim$ 0.1 $M_{\sun}$ (envelope + protostar). 
This amount of material was originally enclosed within a radius of 940 AU in the initial core.
At this 940 AU radius, the initial core has a specific angular momentum of $\sim$ 3.5 $\times$ 10$^{-5}$ km s$^{-1}$ pc. Therefore, the angular momentum which has been carried in during the inside-out collapse is consistent with our upper limit for the angular momentum in the inner region. Moreover, the time scale of the propagation of the expansion wave to this 940 AU radius is $\sim$ 2.2 $\times$ 10$^{4}$ years, on the assumption of the sound speed of 0.2 km s$^{-1}$ \citep{Shu77}. This time scale is smaller than typical Class 0 lifetime (1.7 $\pm$ 0.3 $\times$ 10$^{5}$ yr) \citep{Eno09}.
Hence, the small angular momentum in the inner region in B335 can be explained by the early phase of the inside-out collapse, and the material with  larger angular momenta at the outer part has not yet fallen in and accumulated in the inner region. 

For this scenario, we expect that as the expansion wave propagates outward, a larger amount of the angular momentum will be carried in and accumulated into inner region, 
resulting in a larger radius of the Keplerian disk.
In Table 5, we compare the radius of the (inferred) centrifugally-supported, Keplerian disk around Class 0, I, and II sources. 
The sources associated with the observable large-scale 
Keplerian disks ($r$ $>$ 400 AU),
GG Tau, DM Tau, and LkCa15, 
are in the most evolved stage (Class II). 
On the other hand, the sources in younger evolutional stages (Class I and 0), 
such as L1551 IRS5, L1527, HH 212, and HH211,  
have a smaller inferred radius for the Keplerian disk compared to those around the Class II sources. 
We note that there was no clear direct detection of Keplerian motions around these class 0 and I sources, although some of them show a hint of Keplerian rotation that is indistinguishable from rotation with an $r^{-1}$ dependence \citep{Lom08, Lee09}. 
Nevertheless, jets around these Class 0 and I sources suggests that it is most likely for them to be associated with Keplerian disks (e.g., Shu et al. 1994).
We also note that although HL Tau is classified as a Class II source, it is associated with the infalling material, suggesting that it is actually a Class I source. 
Among these sources, B335 has the smallest inferred Keplerian radius. 
This result may suggest that the Keplerian disk radius could increase 
with evolution of the inside-out collapse scenario \citep{Ter84}.  
We should note that in the latest evolutionary stage (Class II) the increase of the disk radius with evolution could be due to the transportation of the angular momentum inside the disk \citep{Har98, Kit02}.

If the collapse follows Larson-Penston solution \citep{Lar69, Pen69} 
the material at the inner radius with a smaller angular momentum and at the outer radius with a larger angular momentum fall in at almost the same time. 
Hence, we do not expect a smaller specific angular momentum in the central hundred-AU region in B335 unless there is a mechanism to effectively remove the angular momentum of the material coming from the large scale, such as magnetic braking. 
In this case, the difference of the specific angular momentum on a hundred-AU scale between B335 and other protostellar sources could be due to the efficiency of magnetic braking. 
Simulations have shown that angular momentum can be removed effectively during collapse with a strong magnetic field \citep{Bas94, Hen08, Mel08, Mel09}.  
With this Larson-Penston collapse scenario, it would be difficult to explain the increase of Keplerian radius with evolution since the majority of the material falls in at almost the same time. If this is the case, the evolution of the Keplerian radius from Class 0 to Class II could be due to the transportation of  the angular momentum inside the disk \citep{Har98, Kit02}.

\section{Summary}
We have performed detailed imaging and analyses of the SMA Observations of B335 in 1.3 mm continuum, $^{12}$CO (2--1), $^{13}$CO (2--1), and C$^{18}$O (2--1) emission, taken as a part of our large low-mass star-formation project, ``PROSAC''.

1. In $^{12}$CO emission, we found two distinct outflow components associated with B335; one blueshifted ($\Delta$$V$ = -5.8 -- -1.6 km s$^{-1}$) and redshifted ($\Delta$$V$ = 0.6 -- 4.8 km s$^{-1}$) $V$-shaped structure opening towards the east and west of the protostar, respectively, which probably delineates the conical-shaped outflow shell with the eastern side inclined slightly ($\sim$ 10$\degr$) toward us. The other is a compact (1500 AU $\times$ 900 AU), high-velocity blueshifted ($\Delta$$V$ = -37.5 -- -18.5 km s$^{-1}$) 
and redshifted ($\Delta$$V$ = 17.5 -- 36.5 km s$^{-1}$) component to the east and west of the protostar, respectively. The mean propagation velocity ($\sim$ 160 km s$^{-1}$) of the high-velocity components is comparable to the velocity of the associated HH objects (140 -- 170 km s$^{-1}$), and the high-velocity components probably trace molecular jets driven by B335. The $^{13}$CO (2--1) emission traces the low-velocity outflow shell.

2. The C$^{18}$O emission shows a compact ($\sim$ 1500 AU) condensation with an east (blueshifted) to west (redshifted) velocity gradient. Around the systemic velocity ($\Delta$$V$ = 0.0 -- 0.2 km s$^{-1}$) the C$^{18}$O emission shows a similar structure to that in the 1.3 mm continuum emission and elongation perpendicular to the direction of the associated outflow. Hence, the C$^{18}$O emission probably traces the protostellar envelope around B335. The E-W velocity gradient can be interpreted as an infalling gas motion in the flattened disklike envelope, while there is no clear velocity gradient along the N-S direction or sign of the rotation in the envelope. From our simple modeling of the infalling disklike envelope, the central stellar mass and the mass infalling rate were  estimated to be $\sim$ 0.02--0.04 $M_{\sun}$ and $\sim$ 4.8--6.9 $\times$ 10$^{-6}$ $M_{\sun}$ yr$^{-1}$, respectively, and the accretion luminosity was estimated to be $\sim$ 0.7--2.1 $L_{\sun}$. The upper limit of the specific angular momentum is estimated as $\sim$ 7 $\times$ 10$^{-5}$ km s$^{-1}$ pc, which corresponds to a rotational velocity of 0.04 km s$^{-1}$ at a radius of 370 AU.

3. The mass-loss rate ($\sim$ 9.6 $\times$ 10$^{-7}$ $M_{\sun}$ yr$^{-1}$) and the momentum flux ($\sim$ 1.5 $\times$ 10$^{-4}$ $M_{\sun}$ yr$^{-1}$ km s$^{-1}$) of the high-velocity $^{12}$CO jets in B335 are one order of magnitude lower than those in other protostellar sources such as HH 212 and HH 211. There is no thermal SiO emission found in B335, while intense SiO and H$_{2}$ emissions were found in HH 212 and HH 211. These results imply that the jet phenomena in B335 are less active. We suggest that the lower bolometric luminosity and hence the lower mass accretion in B335 are linked to a weaker jet activity than that in the more luminous protostellar sources.
The jet activities in B335 are most likely episodic because of the presence of the chain of the discrete HH objects, and it is considered to be linked to the episodic accretion. The short dynamical time scale of the high-velocity jets ($\sim$ 45 yr) may reflect a recent increase of the mass accretion toward the central protostar.

4. The flattened infalling envelope traced in the C$^{18}$O emission shows no signature of rotation down to a radius of $\sim$ 370 AU. As compared to the previous single-dish studies of the envelope around B335, we found that the specific angular momentum of the envelope rotation decreases from a radius of 20000 AU to 300 AU. 
The upper limit of the specific angular momentum (7 $\times$ 10$^{-5}$ km s$^{-1}$ pc) in the 
region within a radius of 370 AU is one order of magnitude smaller than the angular momentum around other protostellar sources, and the estimated size 
of the centrifugally-supported disk ($\sim$ 6 AU) around B335 is almost two orders of magnitude smaller than that around YSOs in Taurus. 
Other more evolved sources tend to show higher angular momenta in the inner region, and a larger size for the centrifugally-supported disk. We presume that the low specific angular momentum and the small inferred Keplerian disk around B335 could be due to its young age in the course of the inside-out collapse of the core with the increase specific angular momentum as a function of the radius or the more efficient magnetic braking.

\acknowledgments
We are grateful to C.-F. Lee, N. Hirano, and Z.-Y. Li for fruitful discussions. We also acknowledge J. Karr, for checking and improving
English in our manuscrip. We would like to thank all the SMA staff supporting this work. The research of S. T. and N. O. are supported by NSC 97-2112-M-001-003-MY2 and NSC97-2112-M-001-019-MY2, respectively.

\begin{figure}
\epsscale{1}
\plotone{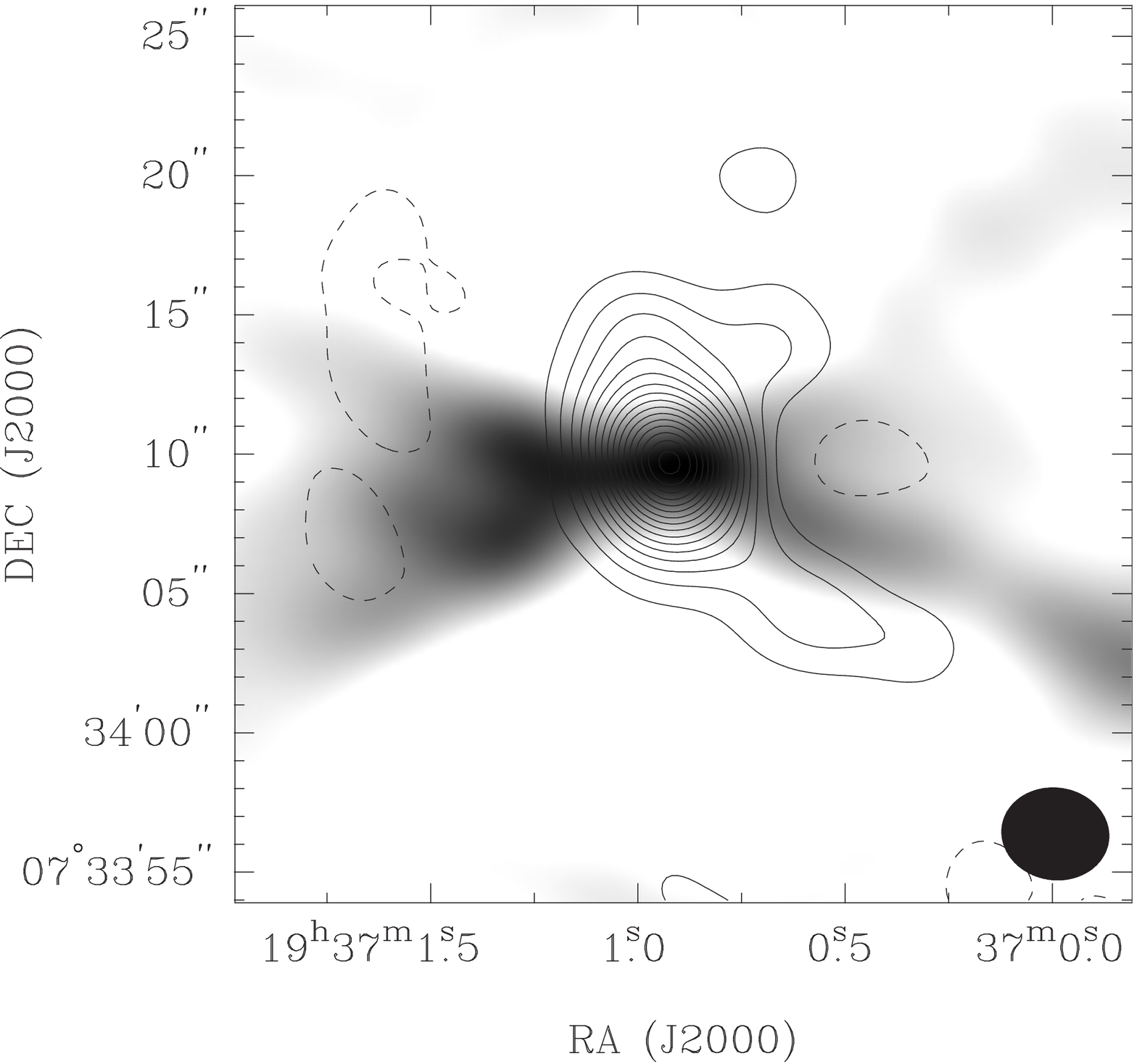}
\caption{1.3 mm continuum map of B335 (contours) overlaid on the $^{12}$CO (2--1) outflow map in linear gray scale whose range is from 6 mJy to 98 mJy. Contour levels are from 3$\sigma$ to 48$\sigma$ in steps of  3$\sigma$, where 1$\sigma$ = 2 mJy. A filled ellipse at the bottom right corner shows the synthesized beam.}
\end{figure}

\begin{figure}
\epsscale{1}
\plotone{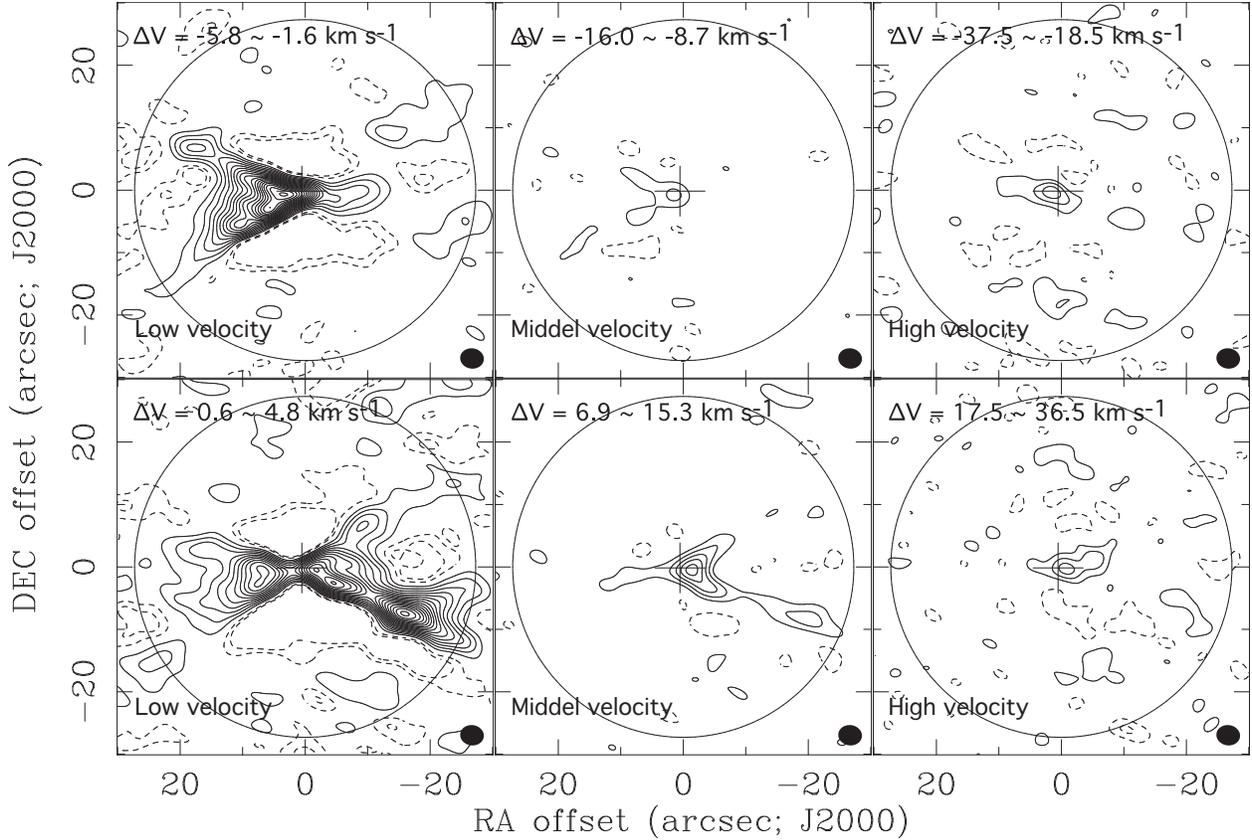}
\caption{Moment 0 maps of the $^{12}$CO (2--1) emission in B335 at different velocity ranges.  For the high velocity, contour levels are from 2$\sigma$ to 6$\sigma$ in steps of 2$\sigma$, where 1$\sigma$ is 1.3 K km s$^{-1}$. For the middle velocity, the contour levels are from 3$\sigma$ to 15$\sigma$ in steps of  3$\sigma$, where 1$\sigma$ is 1.2 K km s$^{-1}$. For the low velocity, the contour levels are from 3$\sigma$ to 71$\sigma$ in steps of 4$\sigma$, where 1$\sigma$ is 1.7 K km s$^{-1}$. Crosses represent the position of the central source, and open circles present the field of view. Filled ellipses at the bottom right corner in each panel show the synthesized beam.}
\end{figure}

\begin{figure}
\epsscale{1}
\plotone{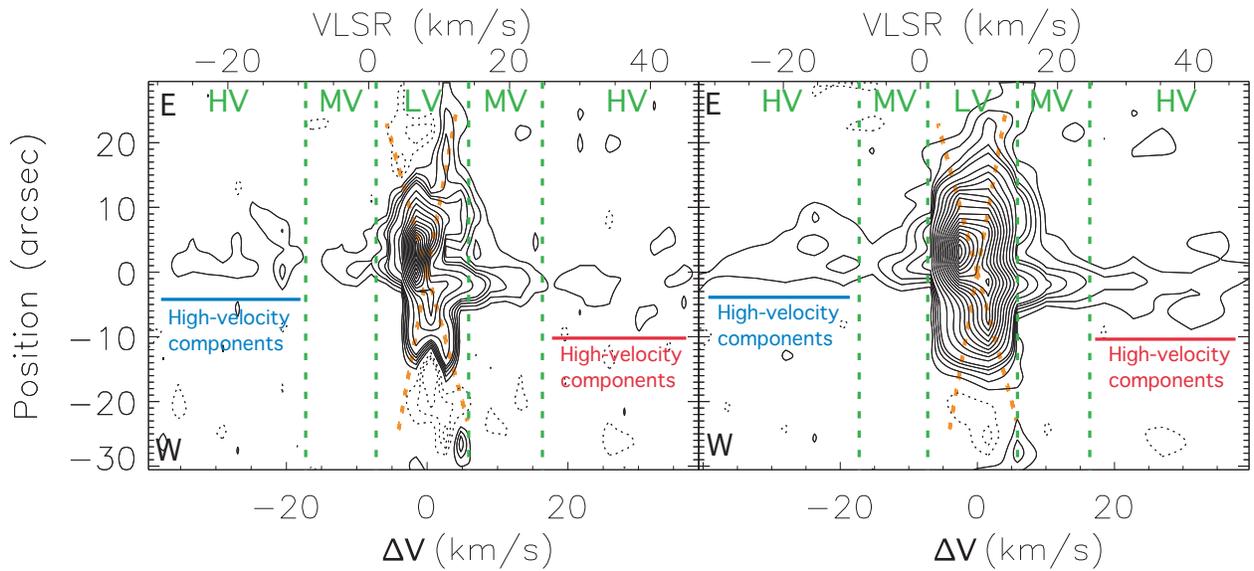}
\caption{P-V diagrams of the $^{12}$CO (2--1) emission along the outflow axis in B335, smoothed over the two ($left$ panel) and four velocity ($right$ panel) channels. Green dotted lines divide the $^{12}$CO emission into three different velocity components, and the blue and red lines show the velocity range of the high-velocity components. Orange dotted crosses represent our simple model of the $^{12}$CO outflow shell. Contour levels are from 2$\sigma$ in steps of 2$\sigma$ until 10$\sigma$, and then in steps of 6$\sigma$, where 1$\sigma$ is 0.2 K in the left panel and 0.1 K in the right panel.}
\end{figure}

\begin{figure}
\epsscale{1}
\plotone{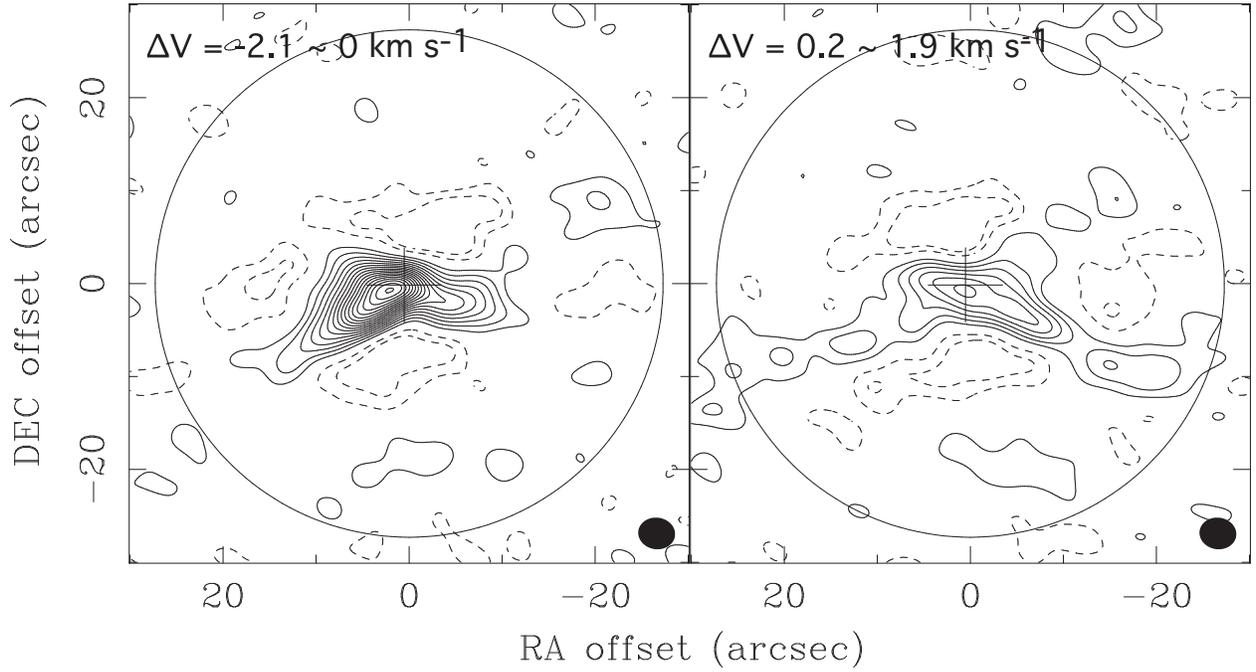}
\caption{Moment 0 maps of the $^{13}$CO (2--1) emission in B335 integrated from $\Delta$$V$ = -2.1 to 0.0 km s$^{-1}$ ($left$ panel) and from $\Delta$$V$ = 0.2 to 1.9 km s$^{-1}$ ($right$ panel) . Contour levels are from 2$\sigma$ in steps of 2$\sigma$, where 1$\sigma$ is 0.3 K km s$^{-1}$. Crosses denote the position of the central source, and open circles represent the field of view. Filled ellipses at the bottom right corner show the synthesized beam.}
\end{figure}

\begin{figure}
\epsscale{1}
\plotone{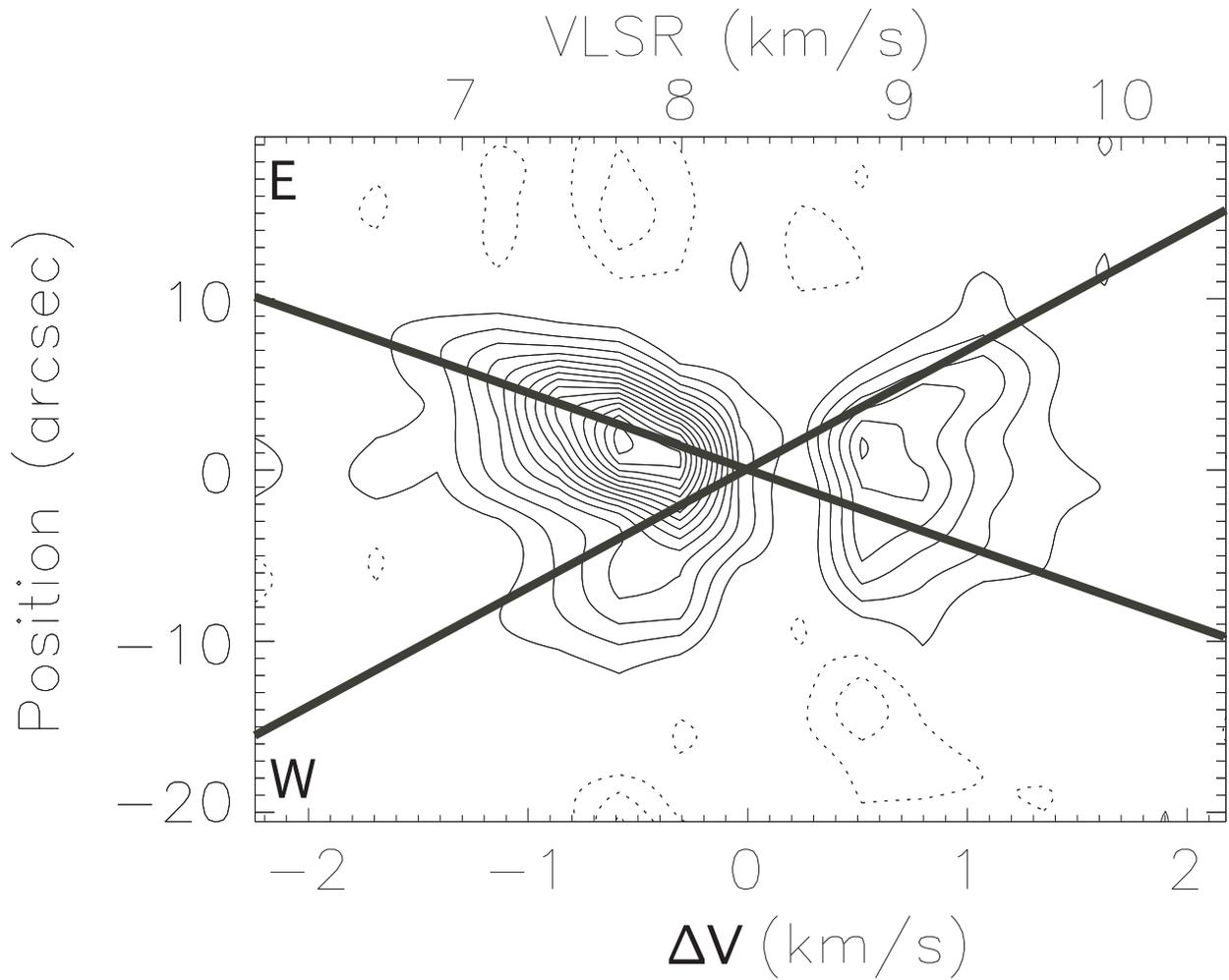} 
\caption{P-V diagram of the $^{13}$CO (2--1) emission along the outflow axis. Bold solid lines show our simple model of the $^{13}$CO outflow shells. Contour levels are from 2$\sigma$ in steps of 2$\sigma$, where 1$\sigma$ is 0.4 K.}
\end{figure}

\begin{figure}
\epsscale{1}
\plotone{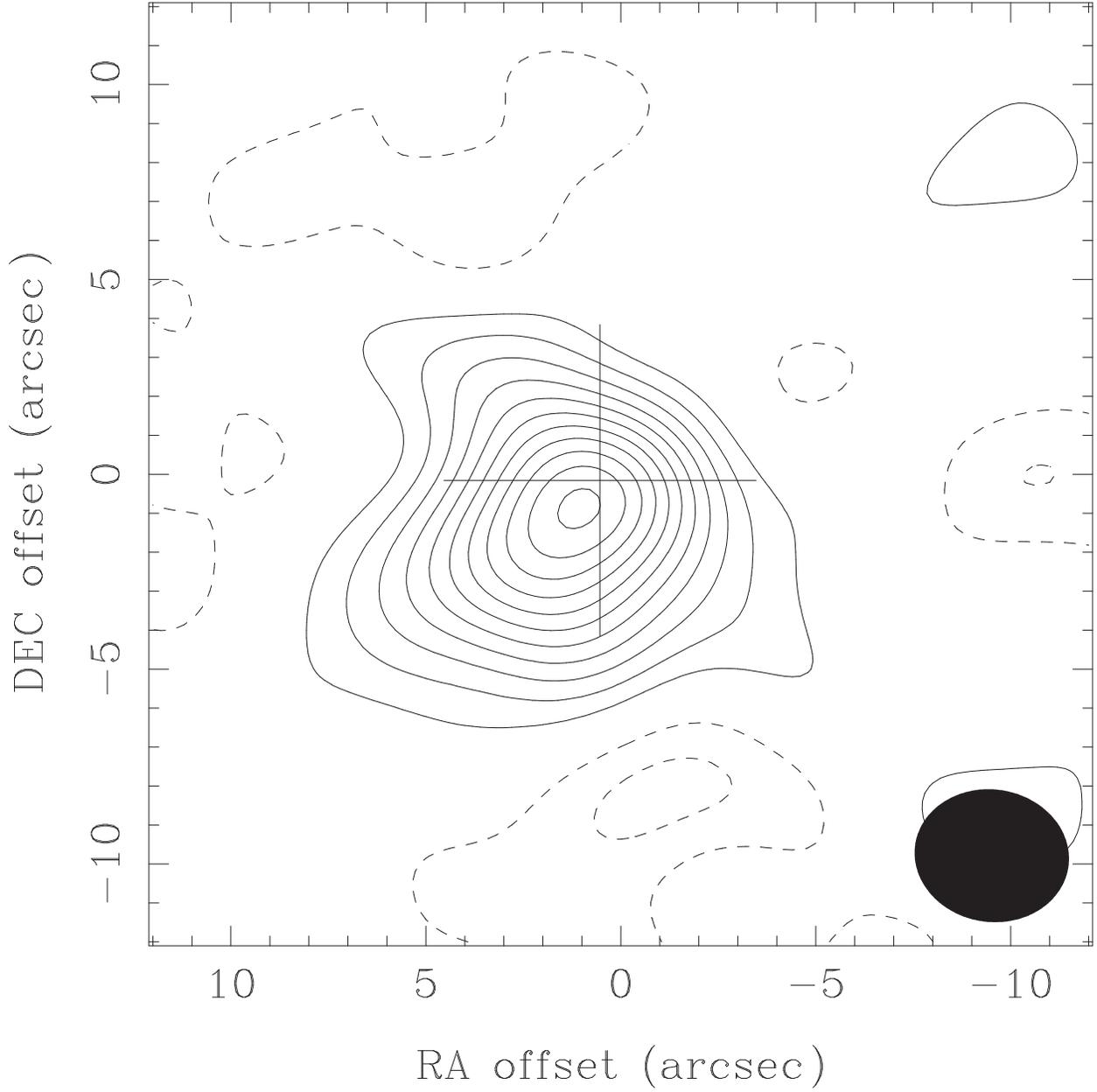} 
\caption{Moment 0 map of the C$^{18}$O (2--1) emission in B335. The integrated velocity range is $\Delta$$V$ = -0.9 -- 0.8 km s$^{-1}$. Contour levels are from 2$\sigma$ to 20$\sigma$ in steps of 2$\sigma$, where 1$\sigma$ is 0.5 K km s$^{-1}$. A cross shows the position of the central source, and a filled ellipse at the bottom right corner shows the synthesized beam.}
\end{figure}

\begin{figure}
\epsscale{1}
\plotone{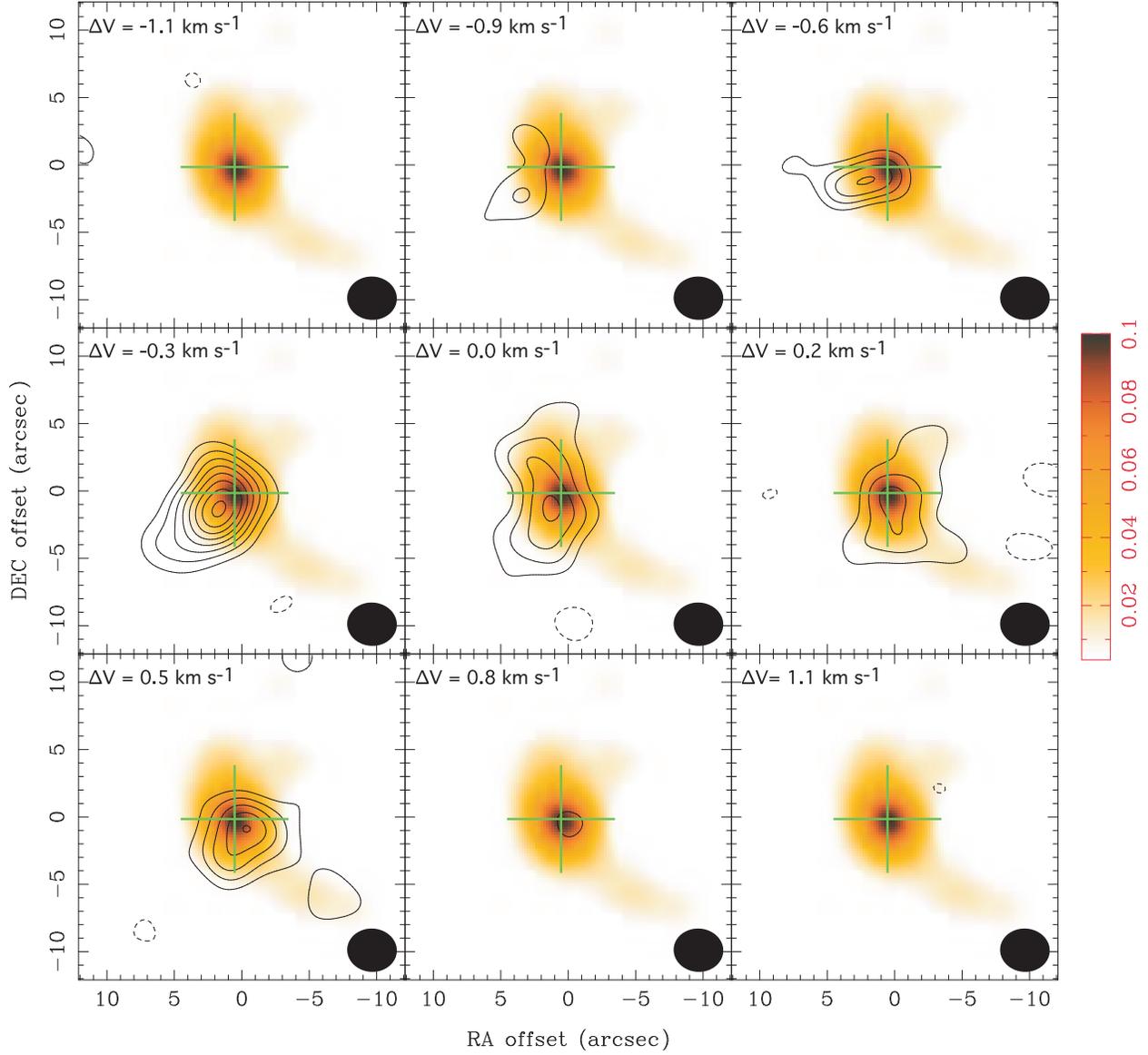}
\caption{Velocity channel maps of the C$^{18}$O (2--1) emission in B335 overlaid on the 1.3 mm continuum emission in gray scale. Contour levels are from 3$\sigma$ in steps of 2$\sigma$, where 1$\sigma$ is 0.6 K. Green crosses show the position of the central source, and filled ellipses at the bottom right corner in each panel show the synthesized beam.}
\end{figure}

\begin{figure}
\epsscale{1}
\plotone{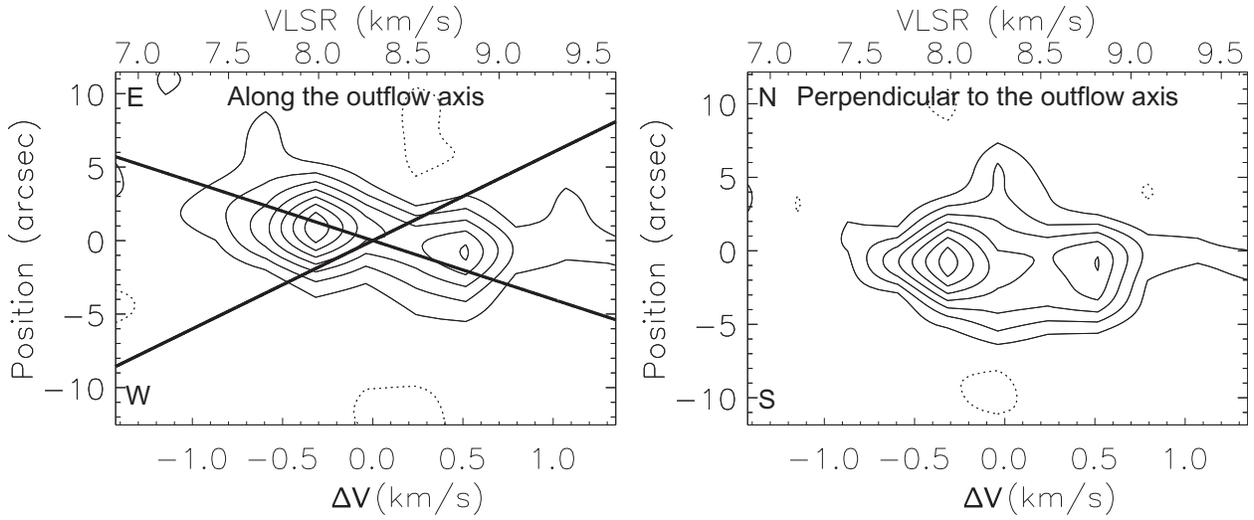} 
\caption{P-V diagrams of the C$^{18}$O (2--1) emission in B335 along the outflow axis (P. A. = 90\degr, $left$ panel) and perpendicular to the outflow axis (P. A. = 0\degr, $right$ panel), passing through the central protostellar position. In the $left$ panel, bold solid lines represent our simple model of the $^{12}$CO outflow shells as shown in Fig. 3. Contour levels are from 2$\sigma$ in steps of 2$\sigma$, where 1$\sigma$ is 0.6 K.}
\end{figure}

\begin{figure}
\epsscale{1}
\plotone{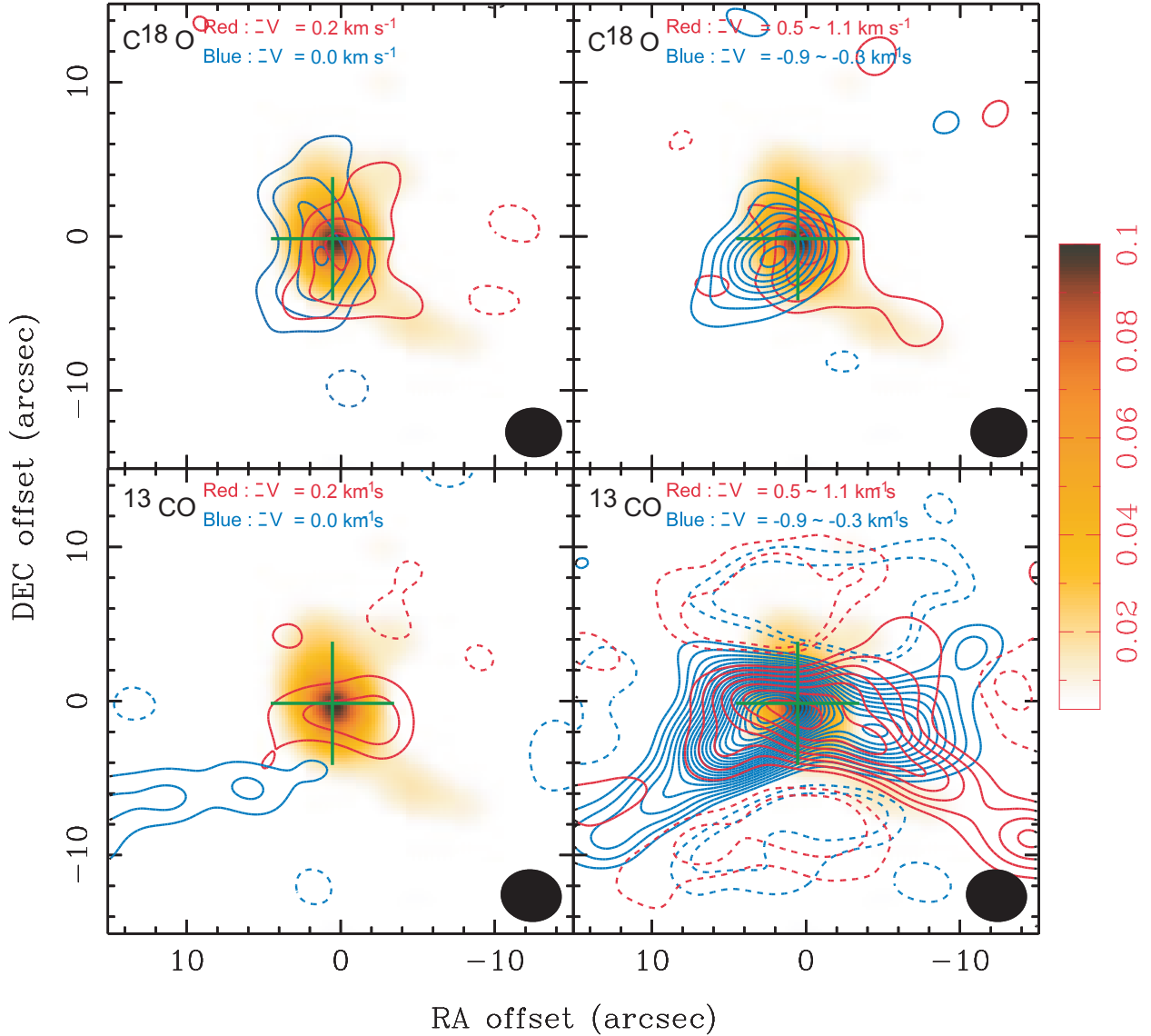}
\caption{Moment 0 maps of the C$^{18}$O (2--1) ($upper$ panels) and $^{13}$CO (2--1) ($lower$ panels) emissions in B335 integrated over four different velocity ranges as written in the Figure, superposed on the 1.3 mm continuum map in gray scale. Contour levels are from 3$\sigma$ in steps of 2$\sigma$, where 1$\sigma$ corresponds to 0.17 K km s$^{-1}$ in the lower-velocity C$^{18}$O map, 0.24 K km s$^{-1}$ in the highly-redshifted C$^{18}$O map, 0.29 K km s$^{-1}$ in the highly-blueshifted C$^{18}$O map, 0.11 K km s$^{-1}$ in the lower-velocity $^{13}$CO map, 0.16 K km s$^{-1}$ in the highly-redshifted $^{13}$CO map, and 0.19 K km s$^{-1}$ in the highly-blueshifted $^{13}$CO map, respectively. Crosses show the position of the central source, and filled ellipses at the bottom right corner in each panel show the synthesized beam of the C$^{18}$O (2--1) ($upper$) and $^{13}$CO (2--1) emissions ($lower$), respectively.}
\end{figure}

\begin{figure}
\epsscale{1}
\plotone{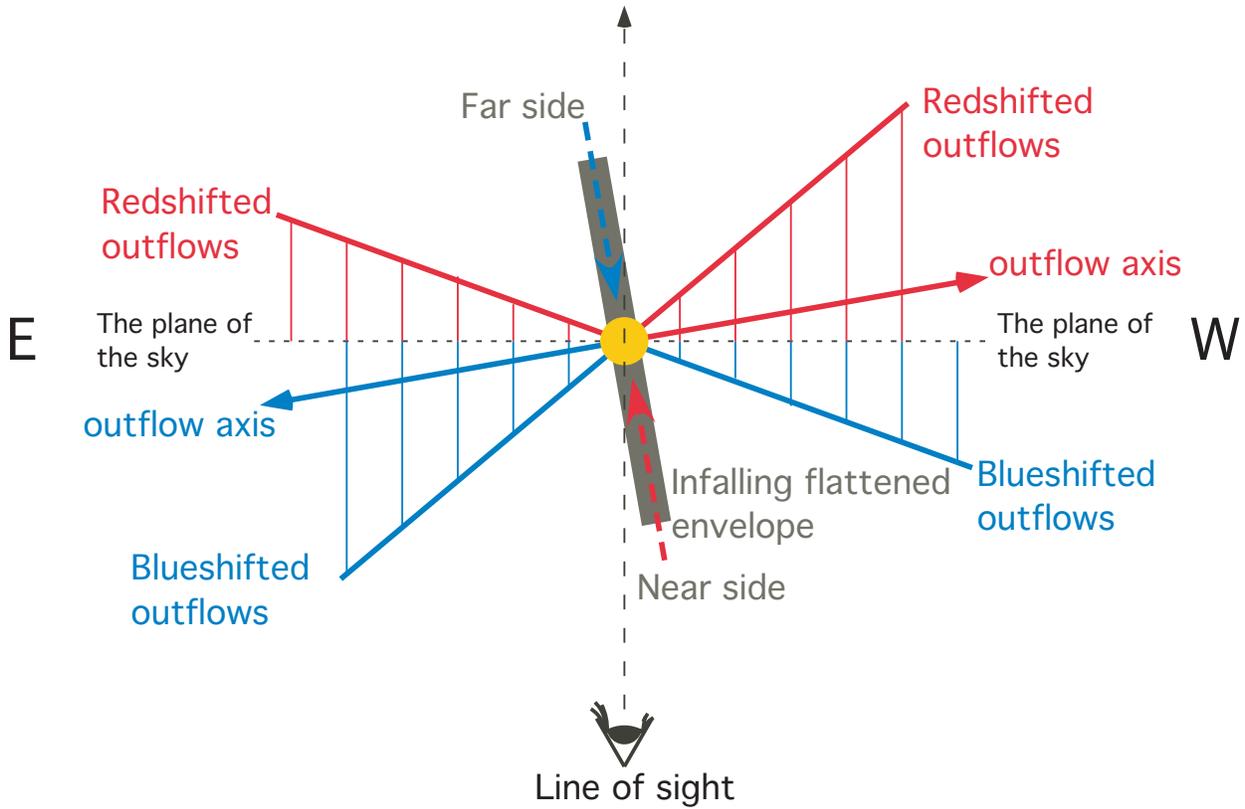}
\caption{Cartoon to show the configuration of the outflow and the infalling flattened envelope viewing from the top. Red and blue solid arrows indicate the redshifted and blueshifted outflow axis, respectively, and red and blue lines show the redshifted and blueshifted outflow. A grey bold line represents the infalling flattened envelop, and red and blue dash arrows on it show the infall motion in the envelope at far side (blueshifted) and near side (redshifted), respectively.}
\end{figure}

\begin{figure}
\includegraphics{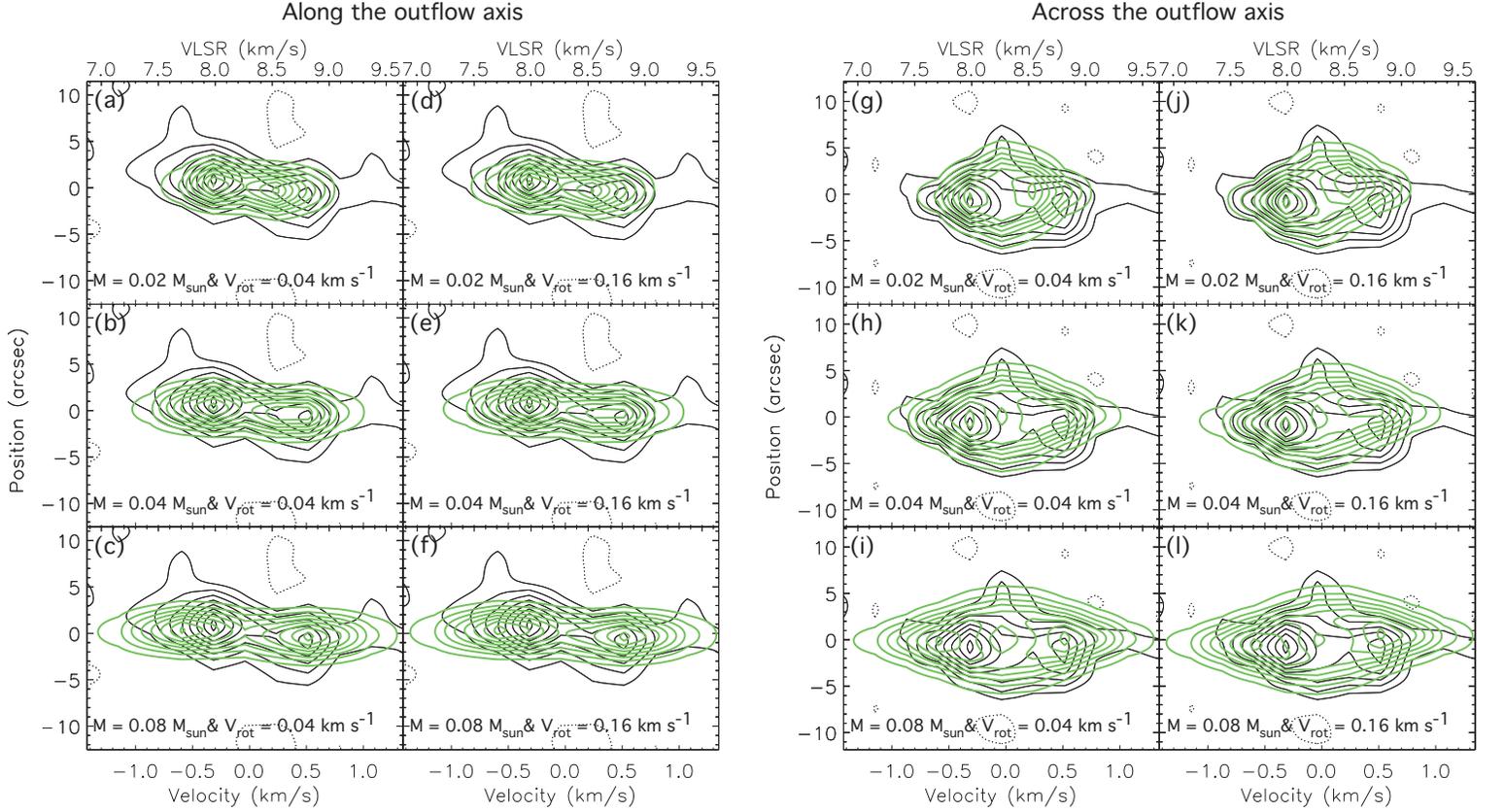}
\vspace{11.0cm}
\caption{P-V diagrams of the C$^{18}$O ($J$ = 2--1) emission in B335 along the disk minor (P. A. = 90\degr; $left$ panel) and major (P. A. = 0\degr; $right$ panel) (black contours) axis, overlaid by our simple model P-Vs of an infalling and rotating gaussian disk (green contours). Green contours in panel $(a)$--$(c)$ and $(g)$--$(i)$ show the model P-V diagrams along and across the outflow axis with a central stellar mass of 0.02, 0.04, and 0.08 $M_{\sun}$ with rotational velocity of 0.04 km s$^{-1}$ at a radius of 370 AU, respectively, and those in panel $(d)$--$(f)$ and $(j)$--$(l)$ show the model P-V diagrams along and across the outflow axis with a central stellar mass of 0.02 and 0.04 and 0.08 $M_{\sun}$ with rotational velocity of 0.16 km s$^{-1}$ at a radius of 370 AU, respectively. Contour levels are from 12\% in steps of 12\% of the maximum intensity, where maximum C$^{18}$O intensity is 9.1 K. }
\end{figure}

\clearpage

\begin{deluxetable}{ccccccc}
\tablewidth{0pt}
\tablecaption{Summary of the observational parameters}
\tablehead{
\colhead{Line} & \colhead{Transition} & \colhead{Beam size (P. A.)} & \colhead{Velocity resolution} & \colhead{Noise level} & \colhead{Weighting} \\
\colhead{} & \colhead{} & \colhead{} & \colhead{(km s$^{-1}$)} & \colhead{(mJy beam$^{-1}$)} & \colhead{}}
\startdata
$^{12}$CO & 2--1 & 3\farcs8\ $\times$ 3\farcs3 (81.7\degr) & 1.06  & 140  & Natural \\
$^{13}$CO & 2--1 & 4\farcs0 $\times$ 3\farcs4 (81.8\degr) & 0.26  & 220  & Natural \\
C$^{18}$O & 2--1 & 3\farcs7 $\times$ 3\farcs2 (86.5\degr) & 0.28  & 280 & Robust = 0.5 \\
\hline\hline
Continuum &  &  & &  & \\
\hline
1.3 mm & \nodata & 3\farcs9 $\times$ 3\farcs3 (81.7\degr) & \nodata & 2 & Natural
\enddata
\end{deluxetable}

\begin{deluxetable}{lccccc}
\tablewidth{0pt}
\tablecaption{Comparison of collimated high-velocity $^{12}$CO emissions among different sources}
\tablehead{
\colhead{Source} & \colhead{Transition} & \colhead{Inclination} & \colhead{Line Width} &\colhead{Velocity\tablenotemark{a}} & \colhead{ref.}\\
 & & \colhead{Angle(\degr)} & \colhead{(km s$^{-1}$)} & \colhead{(km s$^{-1}$)} & 
}
\startdata
 B335 & 2--1 & 10 & 20 & 160 & 1, this paper\\
 HH 212 & 2--1 & 4 & 14 & 120 & 2,3\\
  & 3--2 & & 14 & 190 &4\\
 HH 211 & 3--2 & 5 & 15 & 200 & 5,6 \\
 L1448-mm & 1--0 & 21 & 20 & 200 & 7,8\\
\enddata
\tablerefs{
(1) \cite{Hir88}; (2) \cite{Cla98}; (3) \cite{Lee06}; (4) \cite{Lee07a}; (5) \cite{Lee07b}; (6) \cite{Lee09}; (7) \cite{Bac95}; (8) \cite{Gir01}
}
\tablenotetext{a}{The velocities represent the propagation velocity estimated as $V_{\rm mean} / \sin(\rm inclination\ angle)$.}
\end{deluxetable}

\begin{deluxetable}{lccccccc}
\rotate
\tablewidth{0pt}
\tablecaption{Comparison of $^{12}$CO jet activities among different sources.}
\tablehead{
\colhead{Source} & \colhead{Transition} & \colhead{beam size} & distance & \colhead{$n_{\rm jet}$\tablenotemark{a}} & \colhead{$\dot{M}_{\rm loss}$\tablenotemark{b}} & \colhead{$F$\tablenotemark{c}} & \colhead{ref.}\\
 & & & (pc) & (10$^{5}$ cm$^{-3}$) & (10$^{-6}$ $M_{\sun}$ yr$^{-1}$)& (10$^{-4}$ $M_{\sun}$ yr$^{-1}$ km s$^{-1}$) & 
}
\startdata
 B335 & 2--1 & 3\farcs8 $\times$ 3\farcs4 & 150 & 0.2 & 0.2 & 0.4 & this paper\\
 HH 212 & 2--1 & 2\farcs8 $\times$ 2\farcs3 & 400 & 0.7--1.1 & 0.8--1.3 & 1.0--1.6 & 1\\
  & 3--2 & 1\farcs2 $\times$ 0\farcs8 & & 2.0--2.3 & 3.7--4.3 & 7.0--8.2 & 2\\
 HH 211 & 3--2 & 1\farcs3 $\times$ 0\farcs8 & 280 & 1.6--1.8 & 3.1--3.5 & 6.2--7.0 & 3 \\
\enddata
\tablerefs{(1) \cite{Lee06}; (2) \cite{Lee07a}; (3) \cite{Lee07b}
}
\tablecomments{$n_{\rm jet}$, $\dot{M}_{\rm loss}$, and $F$  are the inferred volume gas density, mass-loss rate, and the momentum flux of the $^{12}$CO jets, respectively. The distance to B335 is $\sim$ 2-3 times smaller than that to HH 211 and HH 212, while the beam size of our observation is $\sim$ 2-3 times larger than that of the other observations. Therefore, our comparison of the jet activities is not affected much by the difference of the linear beam size. }
\tablenotetext{a}{All the volume densities were estimated from the peak intensity of the most inner knot on the assumption of the same $^{12}$CO abundance and the jet diameter of 300 AU.} 
\tablenotetext{b}{The mass-loss rate is derived by $r^{2}\pi V_{\rm jet}n_{\rm jet}\mu$. See $\S$4.1 for details.}
\tablenotetext{c}{The momentum flux is derived by $\dot{M}_{\rm loss} \times V_{\rm jet}$.}
\end{deluxetable}

\begin{deluxetable}{lcccccc}
\tablewidth{0pt}
\tablecaption{Comparison of the protostellar properties between B335 and the other sources associated with SiO jets}
\tablehead{
\colhead{Source} & \colhead{SiO} & \colhead{$M_{*}$} & \colhead{$L_{\rm bol}$} & \colhead{$T_{\rm bol}$} & \colhead{$\frac{L_{\rm bol}}{M_{*}}$} & \colhead{reference} \\
& & ($M_{\sun}$) & ($L_{\sun}$) & (K) & 
}
\startdata
 B335 & no & 0.04 & 1.5 & 37 & 37.5 & 1,2,3, this paper\\
 HH 212 & yes & 0.15 & 14 & 70 & 93.3 & 1,4,5\\ 
 HH 211 & yes & 0.05 & 3.6 & 31 & 72.0 & 6,7,8\\
 L1448-mm & yes & \nodata & 8 & 55 & \nodata & 9,10,11\\
\enddata
\tablecomments{$M_{*}$, $L_{\rm bol}$, and $T_{\rm bol}$ denote the inferred stellar mass, bolometric luminosity, and the bolometric temperature, respectively.}
\tablerefs{
(1) \cite{And00}; (2) \cite{Jor07}; (3) \cite{Stu08}; (4) \cite{Lee06} (5) \cite{Lee07a}; (6) \cite{Lee07b}; (7) \cite{Lee09}; (8) \cite{Fro03}; (9) \cite{Bac95}; (10) \cite{Dut97}; (11) \cite{Mot01}
}
\end{deluxetable}

\begin{deluxetable}{ccccccc}
\tablewidth{0pt}
\tablecaption{Keplerian radius of the protostellar sources}
\tablehead{ 
\colhead{ID} & \colhead{Sources} & \colhead{Class} & \colhead{$j$ (pc km s$^{-1}$)} & \colhead{$r$ (AU)} & \colhead{Keplerian radius (AU)} & \colhead{ref.}}
\startdata
1 & GG Tau & II & 3.5 $\times$ 10$^{-3}$  & 460 & 460 & 1 \\
2 & DM Tau & II & 2.2 $\times$ 10$^{-3}$ & 630 & 630 & 2 \\
3 & HL Tau & II & (6.5--6.8) $\times$ 10$^{-4}$ & 700 & 30-40 & 3,4 \\ 
4 & LkCa15 & II & 2.8 $\times$ 10$^{-3}$ & 430 & 430 & 5 \\
5 & L1527 & I & 4.9 $\times$ 10$^{-4}$ & 2000 & 110 & 6 \\
6 & L1551 IRS 5 & I & (8.2--10.0) $\times$ 10$^{-4}$ & 700--900 & 50--320 & 7,8,9 \\
7 & IRS 63 & I & 8.8 $\times$ 10$^{-4}$ & 100 & 100 & 10 \\
8 & Elias 29 & I & 3.2 $\times$ 10$^{-3}$ & 200 & 200 & 10 \\
9 & HH 212 & 0 & 6.7 $\times$ 10$^{-4}$ & 460 & 70 & 11 \\
10 & HH 211 & 0 & 2.9 $\times$ 10$^{-4}$ & 80 & 80 & 12\\
11 & B335 & 0 & 7 $\times$ 10$^{-5}$ & 370 & 6 & this paper \\
\enddata
\tablecomments{$j$ represents the specific angular momentum derived from disk or envelope rotation at the radius, $r$. If the Keplerian disk has been observed in the sources, the Keplerian radius represents the outer radius of the disk. If the rotating-supported disk has not been found in the source, the Keplerian radius is inferred from the envelope rotation and the estimated stellar mass.}
\tablerefs{(1)\cite{Gui99}; (2) \cite{Gui98}; (3) \cite{Hay93}; (4) \cite{Lin94}; (5) \cite{Qi03}; (6) \cite{Oha97a}; (7) \cite{Mom98}; (8) \cite{Sai96}; (9) \cite{Tak04}; (10) \cite{Lom08}; (11) \cite{Lee06}; (12) \cite{Lee09}; }
\end{deluxetable}

\end{document}